\pdfoutput=1
\newcommand*{\ATLASLATEXPATH}{latex/}
\documentclass[cernpreprint,texlive=2016,txfonts,UKenglish]{latex/atlasdoc}

\usepackage[biblatex=false]{\ATLASLATEXPATH atlaspackage}
\usepackage{\ATLASLATEXPATH atlasbiblatex}

\usepackage{\ATLASLATEXPATH atlasphysics}

\usepackage{cite}

\graphicspath{{logos/}{figures/}}

\usepackage{SbottomPaper2017-defs}
\usepackage{collcell}

\AtlasTitle{Search for supersymmetry in events with \btagged\ jets and missing transverse momentum in $pp$ collisions at $\sqrt{s}=13$~\TeV\ with the ATLAS detector}

\author{The ATLAS Collaboration}

\AtlasRefCode{SUSY-2016-28}

\PreprintIdNumber{CERN-EP-2017-154}

\AtlasJournal{JHEP}

\AtlasAbstract{%
A search for the supersymmetric partners of the Standard Model bottom and top quarks is presented. 
The search uses \intlumi~of $pp$ collision data at $\sqrt{s}=13$~\TeV\ collected by the ATLAS experiment 
at the Large Hadron Collider.  Direct production of pairs of bottom and top squarks 
($\bone$ and $\tone$) is searched for in final states with \btagged\ jets and missing transverse 
momentum.  Distinctive selections are defined with either no charged leptons (electrons or muons) 
in the final state, or one charged lepton. 
The zero-lepton selection targets models in which the $\bone$ is the lightest squark and  decays via $\bone \rightarrow b \neut$, 
where $\neut$~is the lightest neutralino. The one-lepton final state targets models where bottom or top squarks are produced and can decay into 
multiple channels, $\bone \rightarrow b \neut$ and  $\bone \rightarrow t \chipm$, or $\tone \rightarrow t \neut$ and $\tone \rightarrow b \chipm$, 
where $\chipm$ is the lightest chargino and the mass difference $m_{\chipm}- m_{\neut}$ is set to 1~\GeV. 
No excess above the expected Standard Model background is observed. Exclusion limits at 95\% confidence level on the mass of third-generation squarks are derived in various supersymmetry-inspired simplified models.}

\hypersetup{pdftitle={ATLAS document},pdfauthor={The ATLAS Collaboration}}

\RequirePackage[normalem]{ulem} 
\RequirePackage{color}\definecolor{RED}{rgb}{1,0,0}\definecolor{BLUE}{rgb}{0,0,1} 

\begin{document}

\maketitle

\tableofcontents

\section{Introduction}
\label{sec:intro}

Supersymmetry (SUSY)~\cite{Golfand:1971iw,Volkov:1973ix,Wess:1974tw,Wess:1974jb,Ferrara:1974pu,Salam:1974ig} provides an extension 
of the Standard Model (SM) that solves the hierarchy problem~\cite{Sakai:1981gr,Dimopoulos:1981yj,Ibanez:1981yh,Dimopoulos:1981zb} 
by introducing partners of  the known bosons and fermions. In the framework of $R$-parity-conserving models, 
SUSY particles are produced in pairs and the lightest supersymmetric particle (LSP) is stable, providing a possible candidate 
for dark matter~\cite{Goldberg:1983nd,Ellis:1983ew}. In a large variety of models the LSP is the lightest neutralino (\neut). 
Naturalness considerations~\cite{Barbieri:1987fn,deCarlos:1993yy} suggest that the supersymmetric partners of the third-generation 
SM quarks are the lightest coloured supersymmetric particles. This may lead to the lightest bottom squark ($\bone$) and 
top squark ($\tone$) mass eigenstates\footnote{Scalar partners of the left-handed and right-handed chiral components of the bottom 
quark ($\tilde{b}_{\mathrm{L},\mathrm{R}}$) or top quark ($\tilde{t}_{\mathrm{L},\mathrm{R}}$) mix to form mass eigenstates 
for which \bone and \tone are defined as the lighter of the two.} being significantly lighter than the other squarks and the gluinos. 
As a consequence, $\bone$ and $\tone$ could be pair-produced with relatively large cross-sections at the Large Hadron Collider (LHC). 

\begin{figure}[t]
\begin{center}
\subfloat[\label{fig:sbottomdiagram1}]{
\includegraphics[width=0.37\textwidth]{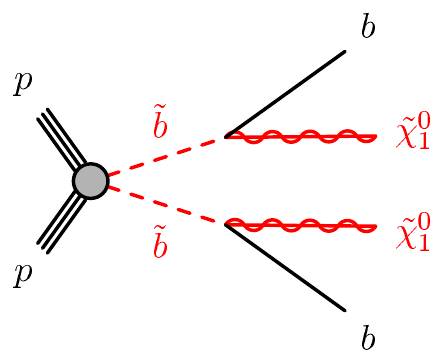}
}
\subfloat[\label{fig:sbottomdiagram3}]{
\includegraphics[width=0.37\textwidth]{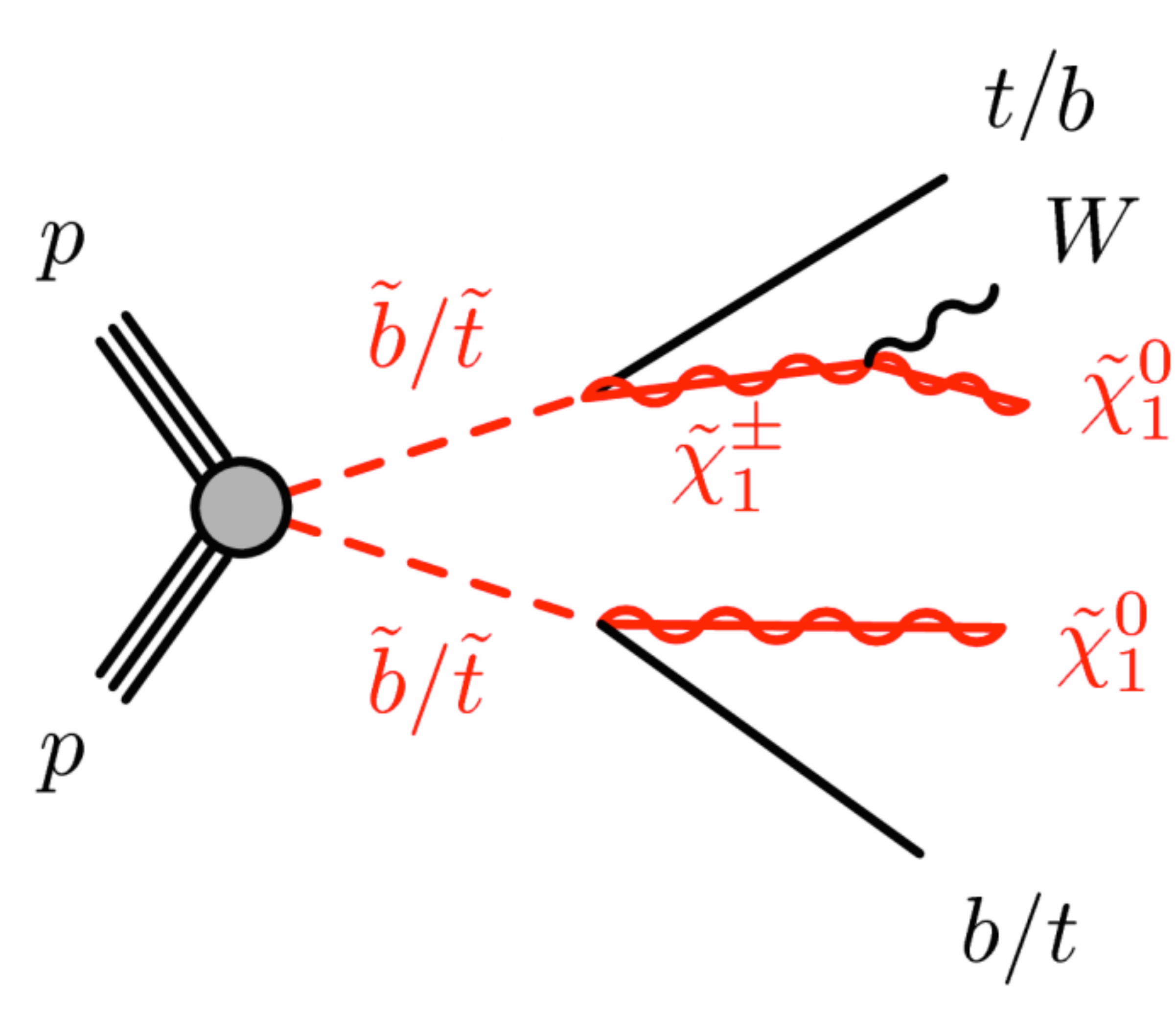}
}
\end{center}
\caption{
Diagrams illustrating the signal scenarios considered for the pair production of bottom and top squarks  
targeted by the (a) zero-lepton and (b) one-lepton channel selections. In (a) bottom squarks 
decay to a bottom quark and the lightest neutralino. In (b), decays via intermediate charginos are kinematically 
available and compete. If the mass difference \dmCN\ is small, the $W$ bosons from chargino decays are off-shell. }

\end{figure}

This paper presents a search for the direct pair production of bottom and top squarks decaying into final states with jets,
 two of them originating from the fragmentation of \bquarks\ (\bjets), and missing transverse 
momentum ($\ptmiss$, whose magnitude is referred to as \etmiss). 
The dataset analysed corresponds to \intlumi~of proton--proton ($pp$) collisions data at $\sqrt{s}=13$~\TeV\ collected by the ATLAS 
experiment during Run 2 of the LHC in 2015 and 2016. 
The third-generation squarks are assumed to decay to the lightest neutralino (LSP) directly or through one intermediate stage. 
The search is based on simplified models inspired by the minimal supersymmetric extension of the SM 
(MSSM)~\cite{Fayet:1976et,Fayet:1977yc,Farrar:1978xj}, where the $\bone$ exclusively 
decays as $\bone \rightarrow b\neut$ or where two decay modes for the bottom (top) squark are allowed and 
direct decays to the LSP, $\bone \rightarrow b\neut$ ($\tone \rightarrow t\neut$) compete with decays 
via an intermediate chargino ($\chipm$) state, $\bone \rightarrow t\chipm$ ($\tone \rightarrow b\chipm$). 
In this case it is assumed that the $\chipm$ is the next-to-lightest supersymmetric particle (NLSP) and is almost degenerate 
with $\neut$, such that other decay products are too low in momentum to be efficiently reconstructed. 
The first set of models lead to final-state events from bottom squark pair production characterized by the 
presence of two \bjets, \etmiss~and no charged leptons ($\ell = e, \mu$), referred to as the zero-lepton channel 
(Figure~\ref{fig:sbottomdiagram1}). 
For mixed decays (direct or through an intermediate stage), the final state of bottom or top squark pair production  
 depends on the branching ratios of the competing decay modes. If the decay modes are equally probable, a large fraction of the signal 
events are characterized by the presence of a top quark, a bottom quark, and neutralinos. 
Hadronic decays of the top quark are targeted by the zero-lepton channel, whilst novel dedicated selections 
requiring one charged lepton, two \bjets and \etmiss~are developed for semi-leptonic decays of the top quark, 
referred to as the one-lepton channel (Figure~\ref{fig:sbottomdiagram3}). 
A statistical combination of the two channels is performed when interpreting 
the results in terms of exclusion limits on the third-generation squark masses.    

Previous searches for the exclusive decay $\bone  \rightarrow b\neut$ with the $\sqrt{s}=13$~\TeV\ LHC Run-2 dataset at ATLAS and CMS have set 
exclusion limits at 95\% confidence level (CL) on $\bone$ masses in such scenarios ~\cite{SUSY-2015-01,Sirunyan:2017kiw}. 
Searches in the context of mixed-decay models were performed only by ATLAS using the Run-1 $\sqrt{s}=8$~\TeV\ dataset and resulted in exclusion limits on the third-generation 
squark mass 
that depend on the branching ratios of the competing decay modes~\cite{SUSY-2014-07}. 

\section{ATLAS detector}
\label{sec:detector}
The ATLAS detector~\cite{PERF-2007-01} is a multi-purpose particle physics detector with a forward-backward symmetric cylindrical
geometry and nearly 4$\pi$ coverage in solid angle.\footnote{ATLAS uses a right-handed coordinate system with its origin at the nominal
interaction point in the centre of the detector. The positive $x$-axis is defined by the direction from the interaction point to the centre
of the LHC ring, with the positive $y$-axis pointing upwards, while the beam direction defines the $z$-axis. Cylindrical coordinates 
$(r,\phi)$ are used in the transverse plane, $\phi$ being the azimuthal angle around the $z$-axis. The pseudorapidity $\eta$ is
defined in terms of the polar angle $\theta$ by $\eta=-\ln\tan(\theta/2)$. Rapidity is defined as $y$ = $0.5 \ln[(E + p_z)/(E-p_z)]$ where $E$ denotes the energy and $p_z$ is the component of the momentum along the beam direction.}
The inner tracking detector consists of pixel and silicon microstrip detectors 
covering the pseudorapidity region $|\eta|<2.5$, surrounded by a transition radiation tracker  
which enhances electron identification in the region $|\eta|<2.0$. 
Between Run 1 and Run 2, a new inner pixel layer, the insertable B-layer~\cite{IBL}, was added at a mean sensor radius of 
3.3~cm. The inner detector is surrounded by a thin superconducting solenoid providing an axial 2~T magnetic field and by a fine-granularity lead/liquid-argon (LAr) electromagnetic calorimeter covering $|\eta|<3.2$. A steel/scintillator-tile calorimeter provides hadronic coverage in the central pseudorapidity range ($|\eta|<1.7$). 
The endcap and forward regions ($1.5<|\eta|<4.9$) of the hadronic calorimeter are made of LAr active layers with either copper or tungsten 
as the absorber material. 
An extensive muon spectrometer with an air-core toroidal magnet system surrounds the calorimeters.
Three layers of high-precision tracking chambers provide coverage in the range $|\eta|<2.7$, while dedicated fast chambers allow triggering in the region $|\eta|<2.4$. The ATLAS trigger system consists of a hardware-based level-1 trigger followed by a software-based high-level trigger~\cite{TRIG-2016-01}.

\section{Data and simulated event samples}
\label{sec:data_mc}
The data used in this analysis were collected by the ATLAS detector in $pp$ collisions at the LHC with a centre-of-mass energy of 13~\TeV\ and a 25~ns proton bunch crossing interval during 2015 and 2016. The full dataset corresponds to an integrated luminosity of 36.1~$\ifb$\ after requiring
that all detector subsystems were operational during data recording. The uncertainty in the combined
2015+2016 integrated luminosity is 3.2\%. It is derived following a methodology similar to that detailed in \Refe~\cite{DAPR-2011-01} 
 from a preliminary calibration of the luminosity scale using $x$--$y$ beam-separation scans performed in August 2015 and May 2016.  
Each event includes on average 13.7 and 24.9 inelastic $pp$ collisions (``pile-up'') in the same bunch crossing in the 2015 and 2016 dataset, respectively.
In the zero-lepton channel, events are required to pass an $\etmiss$\ trigger~\cite{ATL-DAQ-PUB-2016-001}.
 This trigger is fully efficient for events passing the preselection defined in Section~\ref{sec:event_selection}, which requires the offline reconstructed $\etmiss$\ to exceed 200~\GeV. 
Events in the one-lepton channel, as well as events used for control regions, are selected online by a trigger requiring the presence of one electron or muon. The online selection thresholds are such that a plateau of the 
efficiency is reached for charged-lepton transverse momenta of 27~\GeV\ and above.

Monte Carlo (MC) samples of simulated events are used to model the signal and to aid in the estimation of SM background processes, 
except multijet processes, which are estimated from data only. 

All simulated samples were produced using the ATLAS simulation infrastructure~\cite{SOFT-2010-01} using 
GEANT4~\cite{Agostinelli:2002hh}, or a faster simulation~\cite{ATL-PHYS-PUB-2010-013} based on 
a parameterization  of the calorimeter response and GEANT4 for the other detector systems. The simulated events 
are reconstructed with the same algorithm as that used for data. 

SUSY signal samples were generated with \madgraph~\cite{Alwall:2014hca} v2.2.3 at leading order (LO) and interfaced to 
\pythia~v8.186~\cite{Sjostrand:2007gs} with the A14~\cite{ATL-PHYS-PUB-2014-021} set of tuned parameters (tune) for the modelling of the parton showering (PS), 
hadronization and underlying event. The matrix element (ME) calculation was performed at tree level and includes the emission of 
up to two additional partons. The ME--PS matching was done using the CKKW-L~\cite{Lonnblad:2012ix} prescription, 
with a matching scale set to one quarter of the third-generation squark mass.  
The NNPDF23LO~\cite{Ball:2012cx} parton distribution function (PDF) set was used. The cross-sections used to evaluate the signal 
yields are calculated 
to next-to-leading-order (NLO) accuracy in the strong coupling constant, adding the resummation of soft gluon emission at 
next-to-leading-logarithmic accuracy (NLO+NLL)~\cite{Beenakker:1997ut,Beenakker:2010nq,Beenakker:2011fu}.
The nominal cross-section and uncertainty are taken as the midpoint and half-width of an envelope of cross-section predictions using different PDF sets and 
factorization and renormalization scales, as described in \Refe~\cite{Borschensky:2014cia}.

SM background samples were simulated using different MC event generator programs depending on the process. 
The generation of $\ttbar$  was performed by 
the \powhegbox~\cite{Alioli:2010xd} v2 generator with the CT10~\cite{Lai:2010vv} PDF set for the matrix element 
calculations. Single-top-quark events in the $Wt$,  $s$-, and $t-$ channels were generated using the \powhegbox~v1 generator. 
For all processes involving top quarks, top quark spin correlations were preserved. 
The parton shower, fragmentation and the underlying event were simulated using \pythia~v6.428~\cite{Sjostrand:2006za}  with the CTEQ6L1 PDF set and the Perugia 2012~\cite{Skands:2010ak} tune for the underlying event. 
The $h_\mathrm{damp}$ parameter in \powheg, which controls the \pt\ of the first additional emission beyond the Born level and thus regulates the \pt\ of the recoil emission 
against the $\ttbar$ system, was set to the mass of the top quark ($m_t~=~172.5$~\GeV). 
All events with at least one leptonically decaying $W$ boson are retained. 
Fully hadronic $\ttbar$ and single-top events do not contain sufficient \etmiss\ to contribute significantly to the background.
The \ttbar\ samples are normalized using their next-to-NLO (NNLO) cross-section including 
the resummation of soft gluon emission at next-to-NLL accuracy using \textsc{Top++2.0}~\cite{Czakon:2011xx}.
Samples of single-top-quark events are normalized using the NLO cross-sections reported 
in \Refs~\cite{Kidonakis:2011wy,Kidonakis:2010tc,Kidonakis:2010ux} for the $s$-, $t$- and $Wt$-channels, respectively. 

Events containing $W$ or $Z$ bosons with associated jets, including jets from the fragmentation of 
$b$- and $c$-quarks, were simulated using the \sherpa~v2.2.1~\cite{Gleisberg:2008ta} generator.  
Matrix elements were calculated for up to two additional partons at NLO and four partons 
at LO using the \comix~\cite{Gleisberg:2008fv} and \openloops~\cite{Cascioli:2011va} matrix 
element event generators and merged with the \sherpa PS~\cite{Schumann:2007mg} using the ME+PS@NLO prescription~\cite{Hoeche:2012yf}. 
The NNPDF30NNLO~\cite{Ball:2012cx} PDF set was used in conjunction with a dedicated PS tune developed 
by the \sherpa authors. 
Additional \sherpa $Z$+jets samples were produced with similar settings but with up to four partons LO, for the $\gamma$+jets 
studies detailed in Section~\ref{sec:background}. 
The $W$/$Z$+jets events are normalized using their NNLO QCD theoretical cross-sections~\cite{Gavin:2010az}.

Diboson processes were also simulated using the \sherpa generator using the NNPDF30NNLO PDF set in 
conjunction with a dedicated PS tune developed by the \sherpa authors.
They were calculated for up to one ($ZZ$) or zero ($WW, WZ$) additional partons at NLO and up to three additional partons at LO. 
Additional contributions to the SM backgrounds in the signal regions arise from the production of top quark pairs in association 
with $W/Z/h$ bosons and possibly additional jets. 
The production of top quark pairs in association with electroweak vector bosons ($W, Z$) or Higgs bosons 
was modeled by samples generated at NLO using \madgraph~v2.2.3 and showered with \pythia~v8.212. 
Other potential sources of backgrounds, 
such as the production of three or four top quarks or three gauge bosons, are found to be negligible. 

For all samples, except the ones generated using \sherpa, the \evtgen~v1.2.0 program~\cite{EvtGen} was used to simulate the properties of the bottom- and charm-hadron decays.  In-time and out-of-time pile-up interactions from the same or nearby bunch-crossings were simulated by overlaying additional $pp$ collisions generated by \pythia\ v8.186, with the MSTW2008LO~\cite{Martin:2007bv} PDF set, superimposed onto the hard-scattering events to reproduce the observed distribution of the average number of interactions per bunch crossing~\cite{ATL-PHYS-PUB-2012-003}. 

Several samples produced without detector simulation are employed to estimate systematic uncertainties associated with the specific 
configuration of the MC event generators used for the nominal SM background samples. They include variations of the renormalization 
and factorization scales, the CKKW-L matching scale, as well as different PDF sets and fragmentation/hadronization models. 
Details of the MC modelling uncertainties are discussed in \Sect~\ref{sec:systematics}.

\section{Event reconstruction}
\label{sec:event_reco}
The search for pair production of bottom and top squarks is based on two distinct selections of events with \bjets\ and 
large missing transverse momentum, with either no charged leptons  in the final state, 
or requiring exactly one electron or muon (for details, see Section 5). 
For the zero-lepton channel selection, events containing 
charged leptons are explicitly vetoed in the signal and validation regions. Events characterized by the presence 
of exactly one electron or muon with transverse momentum above 27 \GeV\ are retained in the one-lepton selection and 
are also used to define control regions for the zero-lepton channel. Finally, same-flavour opposite-sign (SFOS) 
two-lepton (electron or muon) events with dilepton invariant mass near the $Z$ boson mass are used for 
control regions employed to aid in the estimation of the $Z+$jets background for the zero-lepton channel. 
The details of the reconstruction and selection, as well as the overlap removal procedure 
are given below. 

Selected events are required to have a reconstructed primary vertex consistent with the beamspot 
envelope and to consist of at least two tracks in the inner detector with \pt$>$~0.4~\GeV. 
When more than one such vertex is found, the one with the largest sum of the squares of transverse momenta 
of associated tracks~\cite{ATL-PHYS-PUB-2015-026} is chosen. 

Jet candidates are reconstructed from three-dimensional energy clusters~\cite{PERF-2014-07} in the calorimeter 
using the anti-$k_t$ jet algorithm~\cite{Cacciari:2008gp, Cacciari:2011ma} with a radius parameter of 0.4.
 The reconstructed jets are then calibrated to the particle level by the application of a jet energy scale (JES) derived from $\sqrt{s}=13$~\TeV\ data and simulation~\cite{PERF-2016-04}.
 Quality criteria are imposed to identify jets arising from non-collision sources 
or detector noise, and any event containing such a jet is removed~\cite{ATLAS-CONF-2015-029}. 
Further track-based selections are applied to reject jets with $\pt<60$~\GeV\ and $|\eta|<2.4$ that originate 
from pile-up interactions~\cite{PERF-2014-03}, and the jet momentum is corrected by subtracting the expected 
average energy contribution from pile-up using the jet area method~\cite{ATL-PHYS-PUB-2015-015}. 
Jets are classified as ``baseline'' and ``signal''. Baseline jets are required to have $\pt>20$~\GeV\ and $|\eta|<4.8$. 
Signal jets, selected after resolving overlaps with electrons and muons, are required to 
pass the stricter requirement of $\pt>35$~\GeV\ and $|\eta|<2.8$. 

Jets are identified as \bjets if tagged by a multivariate algorithm which uses information about the impact parameters 
of inner detector tracks matched to the jet, the presence of displaced secondary vertices, and the reconstructed 
flight paths of $b$- and $c$-hadrons inside the jet~\cite{PERF-2012-04}. 
The \btagging working point with a 77\% efficiency, as determined in a sample of simulated \ttbar events, 
was chosen as part of the optimization procedure. The corresponding rejection factors against jets originating 
from \cquarks and from light quarks and gluons at this working point are 6.2 and 134, respectively~\cite{ATL-PHYS-PUB-2016-012}. 
To compensate for differences between data and MC simulation in the \btagging efficiencies and mis-tag rates, correction factors are derived from data and applied to the samples of simulated events~\cite{PERF-2012-04}.  
Candidate \bjets\ are required to have $\pt > 20$~\GeV\ and $|\eta| <$ 2.5.

Electron candidates are reconstructed from energy clusters in the electromagnetic calorimeter matched to a track 
in the  inner detector and are required to satisfy a set of ``loose'' quality 
criteria~\cite{PERF-2016-01,PERF-2013-05,ATL-PHYS-PUB-2015-041}. They are also required to lie within 
the fiducial volume $|\eta|<2.47$. Muon candidates are reconstructed by matching tracks in the inner detector 
with tracks in the muon spectrometer. Events containing one or more muon candidates that have a transverse (longitudinal) 
impact parameter with respect to the primary vertex larger than 0.2~mm (1~mm) are rejected to suppress muons from cosmic rays. 
Muon candidates  are also required to satisfy ``medium'' quality criteria~\cite{PERF-2015-10} and have $|\eta |$ < 2.5. 
All electron and muon candidates must have $\pt>$ 10~\GeV. Lepton candidates remaining after resolving 
overlaps with baseline jets (see next paragraph) are called ``baseline'' leptons. In the control and signal 
regions where lepton identification is required, ``signal'' leptons are chosen from the baseline set 
with \pt$>$~27~\GeV\ to ensure full efficiency of the trigger and are required to be isolated from other 
activity in the detector using a criterion designed to accept at least 95\% of leptons from $Z$ boson decays as detailed in Ref.~\cite{TOPQ-2015-09}.
The angular separation between the lepton and the \bjet\ arising from a semi-leptonically decaying 
top quark narrows as the top quark's \pt increases. This increased collimation is accounted for by varying the radius 
of the isolation cone as max(0.2, 10 \GeV/\ptlepnv), where \ptlepnv is the lepton \pt. 
Signal electrons are further required to satisfy ``tight'' quality criteria. 
Electrons (muons) are matched to the primary vertex by requiring the transverse impact parameter ($d_0$) to 
satisfy $|d_0|/\sigma(d_0)<$~5 (3), and the longitudinal impact parameter ($z_0$) to satisfy $|z_0 \sin \theta|<$~0.5~mm 
for both the electrons and muons. The MC events are corrected to account for differences in the lepton trigger, 
reconstruction and identification efficiencies between data and MC simulation.

The sequence to resolve overlapping electrons, muons and jets begins by removing electron candidates sharing an inner detector track 
with a muon candidate. Next, jet candidates within $\Delta R=\sqrt{(\Delta y)^2 + (\Delta \phi)^2} =0.2$ 
of an electron candidate are discarded, unless the jet is \btagged, in which case the electron is discarded 
since it is likely to originate from a semileptonic \bhadron decay. 
Electrons are discarded if they lie  within $\Delta R = 0.4$  of a jet. Muons 
with \pT below (above) 50~\GeV\ are discarded if they lie within $\Delta R = 0.4$ 
($\Delta R = 0.04 + 10\ \GeV/\pT$) of any remaining jet, except for the case where the number of 
tracks associated with the jet is less than three. 

The missing transverse momentum is defined as the negative vector sum of the \pt of all selected and calibrated physics 
objects (electrons, muons and jets) in the event, with an extra term added to account for soft energy in the event which is not associated 
with any of the selected objects. This soft term is calculated from inner detector tracks with \pt above 0.4~\GeV\ 
matched to the primary vertex to make it more robust against pile-up contamination~\cite{PERF-2014-04,PERF-2011-03}.

Reconstructed photons are not used in the main signal event selections but are selected in the regions employed 
in one of the alternative methods used to estimate the \Zjets\ background, as explained in \Sect~\ref{sec:background}. 
Photon candidates are required to have \pt~$ > 10$~\GeV\ and $|\eta| < 2.37$, 
whilst being outside the transition region $1.37 < |\eta| < 1.52$, to satisfy the tight photon 
shower shape and electron rejection criteria~\cite{PERF-2013-04}. The photons used in this analysis 
are further required to have \pt~$ > 145$~\GeV\  and to be isolated.

\section{Event selection}
\label{sec:event_selection}
Two sets of signal regions (SRs) are defined and optimized to target different third-generation squark decay modes and mass hierarchies of the particles involved.
The zero-lepton channel SRs (b0L) are designed  
to maximize the efficiency to retain bottom-squark pair production events where $\bone \rightarrow b\neut$. The one-lepton channel 
selections (b1L) target SUSY models where bottom squarks decay with a significant branching ratio as $\bone \rightarrow t\chipm$ and 
the lightest chargino is almost degenerate with the lightest neutralino. With these assumptions, the final decay 
products of the off-shell $W$ boson from $\chipm \rightarrow \neut W^{*}$ are too soft to be detected.  
If the branching ratios of the two competing decay modes ($b\neut, t\chipm$) are around 50\%, the final state for the largest 
fraction of signal events is characterized by the presence of a top quark, a bottom quark, and neutralinos escaping the detector. 
Similarly, $\tone$ pair production can lead to an equivalent final state if the $\tone \rightarrow t\neut$ and $\tone \rightarrow b\chipm$ 
decay modes compete. 

\subsection{Discriminating variables}
Several kinematic variables and angular correlations, built from the physics objects defined in the previous section, 
are employed to discriminate SUSY from SM background events and are reported below. 
In the following, signal jets are used and are ordered according to decreasing \pt.
\begin{itemize}
\item \dphimin, \dphiminfour, \dphimintwo : These variables are the minimum $\Delta\phi$ between any of the leading jets 
and the missing transverse momentum vector. The background from multijet processes is characterized by small 
values of this variable. Depending on the signal regions, all, four or two jets are used.
\item $\HT$ : This is defined as the scalar sum of the \pt\ of all jets in the event
\[
\HT = \sum_{i} (p_{\mathrm{T}}^{\mathrm{jet}})_{i},  
\]
where the number of jets involved depends on the signal region. 
In addition, the modified form of $\HT$, referred to as the \htfour\ variable, is used to reject events with 
extra-jet activity in signal regions targeting models characterized by small mass-splitting between the bottom squark 
and the neutralino. In \htfour\ the sum starts with the fourth jet (if any). 
\item $\meff$ : This is defined as the scalar sum of the \pt\ of the jets and the \met, i.e.:
\[
\meff = \sum (p_{\mathrm{T}}^{\mathrm{jet}})_{i} + \met.
\]
The \meff\ observable is correlated with the mass of the pair-produced SUSY particles and 
is employed as a discriminating variable in some of the zero-lepton and one-lepton channel 
selections, as well as in the computation of other composite observables. 
\item $\etmiss/\meff$, $\etmiss/\sqrt{\HT}$ : The first ratio is the $\etmiss$ divided by the  $\meff$, while the second emulates the global \etmiss significance, given that  
the \etmiss resolution scales approximately with the square root of the total hadronic energy in the event. Events with low values for these 
variables are rejected as it is most probable that \etmiss\ arises from jets mismeasurements, caused by instrumental and 
resolution effects.  
\item $\mjj$ : This variable is calculated as the invariant mass of the leading two jets. 
In events where at least one of the leading jets is \btagged, this variable aids in reducing the contamination from \ttbar\ events. 
It is referred to as $\mbb$ for events where the two leading jets are \btagged. 
\item $\mt$ : The event transverse mass \mt is defined as $\mt = \sqrt{2 \ptlepsc \etmiss - 2 \ptlepvec \cdot \ptmissvec} $ 
and is used in the one-lepton control and signal regions to reduce the \Wjets\ and \ttbar~backgrounds.   
\item $\mlbmin$ : The minimum invariant mass of the lepton and one of the two \bjets\ is defined as: 
\[
\mlbmin = \textrm{min}_{i = 1,2}\left(m_{\ell b_i}\right). 
\]
This variable is bound from above by $\sqrt{\mtop^2 - m_W^2}$ for $\ttbar$ production, and it is used to distinguish $\ttbar$  
 contributions from $Wt$-channel single-top-quark events in the one-lepton control regions. 
\item Contransverse mass (\mct) \cite{Tovey:2008ui}: This is the main discriminating variable in most of 
the zero-lepton channel signal regions~\cite{SUSY-2013-05}. It is used to measure the masses of pair-produced semi-invisibly 
decaying heavy particles. For identical decays of two heavy particles (e.g. the bottom squarks decaying exclusively  
as $\bone \rightarrow b\neut$) into two visible particles $v_{1}$ and $v_{2}$ (the $b$-quarks), and two invisible 
particles $X_1$ and $X_2$ (the $\neut$ for the signal), \mct\ is defined as 
\[
\mct^{2}(v_1, \ v_2) = [\et(v_1)+\et(v_2)]^{2} - [\mathbf{p}_{\mathrm T}(v_1) - \mathbf{p}_{\mathrm T}(v_2)]^{2}, 
\]
\noindent with $\et = \sqrt{\pt^{2}+m^{2}}$, and it has a kinematical endpoint at $\mct^{\mathrm{max}} = (m_{i}^2 - m_{X}^2)/m_{i}$
where $i$ is the initially pair-produced particle. This variable is extremely effective in suppressing the top-quark pair 
production background ($i = t, X=W$), for which the endpoint is at 135~\GeV. 
\item $\minmtfour$ : This is the minimum of the transverse masses calculated using any of the leading four jets and the \met in the event. 
For signal scenarios with low values of $\mct^{\mathrm{max}}$, this kinematic variable is an alternative discriminating  
variable to reduce the \ttbar\ background. 
\item $\amtTwo$ :  The asymmetric transverse mass \cite{Barr:2009jv,Konar:2009qr} is a kinematic variable which can be used 
to separate processes in which two decays giving missing transverse momentum occur, and it is the main discriminating observable 
in the one-lepton channel signal regions. The \amtTwo\ definition is based on the stransverse mass (\mttwo)~\cite{Lester:1999tx}:
  \begin{equation}
   \mttwo^{2} ( \chi ) = \underset{ {\mathbf q}_{\mathrm T}^{(1)}   +   {\mathbf q}_{\mathrm T}^{(2)} =  {\mathbf p}_{\mathrm T}^{\mathrm miss}    }{\min} \left [ \max \left \{  \trmass^{2}( {\mathbf p}_{\mathrm T}(v_{1}) , \xspace {\mathbf q}_{\mathrm T}^{(1)} ; \chi) , \trmass^{2} ( {\mathbf p}_{\mathrm T}(v_{2}) , {\mathbf q}_{\mathrm T}^{(2)} ; \chi)  \right \}  \right ], \nonumber
  \end{equation}
  \noindent where ${\mathbf p}_{\mathrm T}(v_{i})$ are reconstructed transverse momenta vectors and   
${\mathbf q}_{\mathrm T}^{(i)}$ represent the missing transverse momenta from the two decays, with a total missing transverse 
momentum, $\ptmiss$; $\chi$ is a free parameter representing the unknown mass of the invisible 
particles -- here assumed to be zero. The \textit{a} in \amtTwo\ indicates that the two visible decay legs are asymmetric, i.e.\ 
not composed of the same particles. 

In the case of events with one lepton (electron or muon) and two \bjets, the \mttwo\ variable is calculated for different values of $ {\mathbf p}_{\mathrm T}(v_{1})$ and ${\mathbf p}_{\mathrm T}(v_{2})$,  
by grouping the lepton and the two \bjets\ into two visible objects $v_1$ and $v_2$. 
The lepton needs to be paired with one of the  two \bjets\ and the choice is driven by the value of $\mlb(n)$ -- the invariant mass of the $n^{\mathrm{th}}$ \btagged jet and the lepton. 
If the two particles are correctly associated, this value has an upper bound given by the top quark mass. 
The value of \amtTwo\ is thus computed accordingly:
  \begin{itemize}
  \item If $\mlb(1)$ and $\mlb(2)$ are both $> 170$~\GeV, neither of the two associations is compatible with the \bjet\ 
and the lepton originating from a top decay, so the event is rejected since all control, validation and signal regions require the smaller value of $\mlb$ to be  $< 170$~\GeV.
  \item If $\mlb(1)$ is $< 170$~\GeV\ and $\mlb(2)$ is $> 170$~\GeV, \amtTwo\ is calculated with $ v_{1} = b_{1} + \ell $ 
and $ v_{2} = b_{2}$. This is done because only the first pairing is compatible with a top quark decay.
  \item Similarly, if $\mlb(1)$ is $> 170$~\GeV\ and $\mlb(2)$ is $< 170$~\GeV, \amtTwo\ is calculated with $ v_{1} = b_{1} $ and $ v_{2} = b_{2} + \ell$. 
  \item If $\mlb(1)$ and $\mlb(2)$ are both $< 170$~\GeV, \amtTwo\ is calculated in both configurations and 
its value is taken to be the smaller of the two. This must be done because, according to the $\mlb$ check, 
both pairings would be acceptable.
  \end{itemize}
\item ${\cal A}$ : This is the \pt\ asymmetry of the leading two jets and is defined as:
\[
    {\cal A} = \frac{\pt(j_1) - \pt(j_2)}{\pt(j_1) + \pt(j_2)}. 
\]
The ${\cal A}$ variable is employed in scenarios where the mass-splitting between the bottom squark 
and the neutralino is small ($<20$~\GeV) and the selection exploits the presence of a high-momentum jet from initial-state radiation (ISR). 
\end{itemize}
\subsection{Zero-lepton channel selections}
The selection criteria for the zero-lepton channel SRs are summarized in Table~\ref{tab:SR0l} 
and have the main requirement of no baseline leptons with $\pt>10$~\GeV\ and two \btagged\ jets. To exploit the 
kinematic properties over the large range of $\bone$ and $\neut$ masses explored, three sets of 
SRs are defined. 

The b0L-SRA regions are optimized to be sensitive to models with large mass-splitting between the \bone and 
the \neut, $ \dmBN > 250$~\GeV. Incremental thresholds are imposed on the main discriminating 
variable, \mct, resulting in three overlapping regions (\mct$>$350, 450 and 550~\GeV). 
Only events with \etmiss$>250$~\GeV\ are retained to ensure full efficiency of the  trigger and comply with 
the expected signal topology. The two leading jets are required to be \btagged whilst contamination from backgrounds with high jet multiplicity, 
particularly \ttbar production, is suppressed by vetoing events with a fourth jet with $\pt>50$~\gev. 
To discriminate against multijet background, events where \etmiss is aligned with a jet in the 
transverse plane are rejected by requiring $\dphiminfour>0.4$, and  $\etmiss/\meff>0.25$. 
A selection on the invariant mass of the two \bjets ($\mbb>200$~\GeV) is applied to further enhance 
the signal yield over the SM background contributions. 

The b0L-SRB region targets intermediate mass-splitting between \bone and \neut, $50<\dmBN<250$~\GeV. 
In these scenarios, the selections based on the \mct\ and \mbb\ variables are no longer effective and the variable 
\minmtfour\ is employed to reduce SM background contributions from \ttbar\ production, with 
events selected if $\minmtfour>250$~\GeV. 
No more than four signal jets are allowed, to reduce additional hadronic activity in the 
selected events. 
As opposed to the b0L-SRA criteria, no veto based on the fourth jet \pt is applied. 
A series of selections on the azimuthal angle between the two \btagged jets and the \met are implemented 
($|\Delta\phi(b_1,\met)| < 2.0$ and $|\Delta\phi(b_2,\met)| < 2.5$) to reduce \Zjets background events. 

Finally, the b0L-SRC region targets events where a bottom squark pair is produced in association 
with a jet from ISR. This  selection provides sensitivity to 
models with a small mass difference between the \bone\ and the \neut, $\dmBN < 50$~\GeV, such 
that a boosted bottom squark pair would satisfy the trigger requirements.  
To efficiently suppress $t\bar{t}$ and $W$+jets backgrounds, events are selected with one 
high-\pt non-\btagged\ jet and \etmiss$>500$~\GeV\ such that 
$\Delta\phi(j_1,\met) > 2.5$. 
Stringent requirements on the minimum azimuthal angle between the jets and \etmiss\ 
are not suited for these scenarios where \bjets have softer momenta and are possibly aligned with \met. 
A large asymmetry ${\cal A}$ is required to reduce the multijet background 
while loosening the selection on the minimum azimuthal angle between the jets and \etmiss~to  
\dphimintwo~$>$0.2, and relaxing the \pt\ threshold on signal jets to 20~\GeV.  
\begin{table}[!htpb]
  \centering

  \caption{Summary of the event selection in each signal region for the zero-lepton channel. 
For SRA, the ``x'' denotes the \mct\ selection used. 
The term lepton is used in the table to refer to baseline electrons and muons. Jets ($j_1$, $j_2$, $j_3$, $j_4$ and $j_5$) 
are labelled with an index corresponding to their decreasing order in \pt.} 
\vspace{0.2cm}
  \label{tab:SR0l}
    \begin{tabular*}{\textwidth}{@{\extracolsep{\fill}} lc@{} c c }
    \hline 
    \toprule
          & b0L-SRAx &   b0L-SRB &  b0L-SRC \\
    \midrule
    Lepton veto & \multicolumn{3}{c}{No $e/\mu$ with \pt $>$10~\GeV\ after overlap removal} \\ 
    $N_{\textup{jets}}$ $(p_\mathrm{T}>35~\GeV)$ &   \multicolumn{1}{c}{2--4}    & \multicolumn{1}{c}{2--4}     & \multicolumn{1}{c}{-}    \\ 
    $N_{\textup{jets}}$ $(p_\mathrm{T}>20~\GeV)$ &   \multicolumn{1}{c}{-}      & \multicolumn{1}{c}{-}       & \multicolumn{1}{c}{2--5}   \\ 
    $\pt(j_1)$ [\GeV]       &  \multicolumn{1}{c}{$>130$}      & \multicolumn{1}{c}{$>50$}   & \multicolumn{1}{c}{$>500$}  \\ 
    $\pt(j_2)$ [\GeV]    &  \multicolumn{1}{c}{$>50$}       & \multicolumn{1}{c}{$>50$}   & \multicolumn{1}{c}{$>20$}   \\ 
    $\pt(j_4)$ [\GeV]          &  \multicolumn{1}{c}{$<50$}       & \multicolumn{1}{c}{-}       & \multicolumn{1}{c}{-}    \\ 
    $\htfour$ [\GeV]             &  \multicolumn{1}{c}{-}           & \multicolumn{1}{c}{-}       & \multicolumn{1}{c}{$<70$}  \\ 
    \bjets                      &  $j_1$ and $j_2$               & any 2                         & $j_2$ and ($j_3$ or $j_4$ or $j_5$) \\
    \met  [\GeV]                 &  \multicolumn{1}{c}{$>250$}      & \multicolumn{1}{c}{$>250$}  & \multicolumn{1}{c}{$>500$} \\ 
    \met/\meff                  &  \multicolumn{1}{c}{$>0.25$}      & \multicolumn{1}{c}{-}       & \multicolumn{1}{c}{-}            \\ 
    \dphiminfour                &  \multicolumn{1}{c}{$>0.4$}       & \multicolumn{1}{c}{$>0.4$}  &\multicolumn{1}{c}{-}            \\ 
    \dphimintwo                 &  \multicolumn{1}{c}{-}            & \multicolumn{1}{c}{-}       & \multicolumn{1}{c}{$>0.2$}   \\ 
    $\Delta\phi(b_1,\met)$      & \multicolumn{1}{c}{-}             & $<2.0$         & \multicolumn{1}{c}{-}    \\ 
    $\Delta\phi(b_2,\met)$      & \multicolumn{1}{c}{-}             & $<2.5$         & \multicolumn{1}{c}{-}   \\ 
    $\Delta\phi(j_1,\met)$      & \multicolumn{1}{c}{-}             & \multicolumn{1}{c}{-}       &\multicolumn{1}{c}{$>2.5$}    \\ 
    \mjj  [GeV]                 &  \multicolumn{1}{c}{$>200$}      & \multicolumn{1}{c}{-}       &     \multicolumn{1}{c}{$>200$}    \\ 
    \mct  [\GeV]                 & $>$350, 450, 550  & \multicolumn{1}{c}{-}  & \multicolumn{1}{c}{-}  \\ 
    $\minmtfour$ [\GeV]   &  \multicolumn{1}{c}{-}  & $>250$ & \multicolumn{1}{c}{-} \\ 
    \meff\ [\GeV]              &  \multicolumn{1}{c}{-}           & \multicolumn{1}{c}{-}       & \multicolumn{1}{c}{$>1300$}   \\ 
    ${\cal A}$                         &  \multicolumn{1}{c}{-}            &      \multicolumn{1}{c}{-}                & $>0.8$   \\ 
    \hline
  \end{tabular*}
\end{table}

\subsection{One-lepton channel selections}
The selection criteria for the one-lepton channel SRs are summarized in Table~\ref{tab:SR1L}. 
Events are required to have exactly one signal electron or muon and no additional baseline leptons, 
two \btagged jets and a large \etmiss. 
Similarly to the zero-lepton channel, three sets of SRs are defined to maximize the sensitivity 
depending on the mass hierarchy between $\bone$($\tone$) and $\chipm \approx \neut$. 

The b1L-SRA regions are optimized for models with large \dmBN: events 
are required to have large \met and $\met/\sqrt{\HT}$ and \dphimin~above 
0.4 to reduce the multijet background contributions to negligible levels. Requirements on the \mt\ and \amtTwo\ variables 
to be above 140~\GeV\ and 250~\GeV, respectively, are set to reject \Wjets\ and \ttbar\ events whilst   
the selection on the invariant mass of the two \bjets ($\mbb>200$~\GeV) is applied to further enhance
the signal yield over the SM background contributions. 
Two incremental thresholds are finally imposed on \meff\ (600 and 750~\GeV) to define two
overlapping signal regions. 

The b1L-SRB region is designed to be sensitive to compressed mass spectra, hence  low $\mbb$ is expected, and the 
selections on the \mt\ and \amtTwo\ variables must be relaxed to 
avoid loss of signal events. The \mtbmin\ is employed to discriminate signal from \ttbar\ events, 
which is the dominant SM background contribution. 

A third region, referred to as b1L-SRA300-2j, is defined similarly to the b1L-SRAs but requiring no extra 
jets beside the two \bjets~and \meff~above 300~\GeV. Such a selection also targets SUSY models 
characterized by compressed mass spectra. It is kinematically similar to the signal region in the 
Run-1 analysis~\cite{SUSY-2014-07} with a veto requirement on the number of jets with \pt $>50$~\GeV.  
 
\begin{table}[htbp!]
  \centering
  \caption{Summary of the event selection in each signal region for the one-lepton channel. 
For SRA, the ``x'' denotes the \meff\ selection used. 
The term lepton is used in the table to refer to signal electrons and muons. 
Jets ($j_1$, $j_2$) are labelled with an index corresponding to their decreasing order in \pt.} 
\vspace{0.2cm}
  \label{tab:SR1L}
    \begin{tabular*}{\textwidth}{@{\extracolsep{\fill}} lc@{} c c }
      \toprule
           & b1L-SRAx &   b1L-SRA300-2j & b1L-SRB \\
      \midrule
      Number of leptons ($e,\mu$) & \multicolumn{1}{c}{1} & \multicolumn{1}{c}{1} & \multicolumn{1}{c}{1} \\
      $N_{\mathrm{jets}}$ $(p_\mathrm{T}>35~\GeV)$ &  \multicolumn{1}{c}{$\geq 2$ } & \multicolumn{1}{c}{ = 2 } &  \multicolumn{1}{c}{$\geq 2$ } \\
      \bjets   & any 2 & $j_1$ and $j_2$ & any 2 \\
      \met [\GeV]   &  \multicolumn{1}{c}{ $> 200$ }  & \multicolumn{1}{c}{ $> 200$ }  & \multicolumn{1}{c}{ $> 200$ } \\
      $\met/\sqrt{\HT}$ [\GeV$^{1/2}$] &  \multicolumn{1}{c}{ $> 8$ } & \multicolumn{1}{c}{$> 8$ } & \multicolumn{1}{c}{$> 8$ } \\
      \mlbmin [\GeV] & $ < 170$  & $ < 170$ & $ < 170$  \\ 
      \dphimin  &  \multicolumn{1}{c}{ $> 0.4$ }  & \multicolumn{1}{c}{ -- } &\multicolumn{1}{c}{ $> 0.4$ }  \\
      \dphimintwo  &  \multicolumn{1}{c}{ -- }  & \multicolumn{1}{c}{ $>0.4$ } &\multicolumn{1}{c}{ -- }  \\
      \amtTwo\ [\GeV]  & $ > 250$ & $ > 250$ & $ > 200$   \\
      \mT [\GeV] & $ > 140$ & $ > 140$ & $ > 120$   \\
      \mbb\ [\GeV] & $ > 200$  & $ > 200$ & $ < 200$ \\
      \meff\ [\GeV] &  \multicolumn{1}{c}{ $> 600, 750$ } & $ > 300$ & $ > 300$ \\
      $\minmtb$ [\GeV] & -- & -- & $> 200$ \\
      $\Delta\phi(b_1,\met)$ &    -- & -- & $>2.0$ \\
      \bottomrule
    \end{tabular*}

\end{table}

\section{Background estimation}
\label{sec:background}
Monte Carlo simulation is used to estimate the background yield in the signal regions. The MC prediction for the major backgrounds is  normalized to data in control regions (CR) constructed to enhance a particular background and to be kinematically similar but mutually exclusive to the signal regions. The control regions are defined by explicitly requiring the presence
of one or two leptons (electrons or muons) in the final state together with further selection criteria similar
to those of the corresponding signal region. To ensure that the b0L and b1L analyses can be statistically combined, the CRs associated with b0L and b1L SRs are mutually exclusive, with the exception of the single-top CR, where the same CR is used for both channels. 

The expected SM backgrounds are determined separately for each SR with a profile likelihood fit~\cite{Baak:2014wma}, referred to as the background-only fit. The fit uses as a constraint the observed event yields in a set of associated CRs to adjust the  normalization of the main backgrounds, assuming that no signal is present. 
The inputs to the fit for each SR include the number of events observed in its associated CRs and the number of events predicted by simulation in each region for all background processes. The latter are described by Poisson statistics. The systematic uncertainties in the expected values are included in the fit as nuisance parameters. They are constrained by Gaussian distributions with widths corresponding to the sizes of the uncertainties and are treated as correlated, when appropriate, between the various regions. The product of the various probability density functions forms the likelihood, which the fit maximizes by adjusting the background normalization and the nuisance parameters. Finally, the reliability of the MC extrapolation of the SM background estimate outside of the control regions is evaluated in several validation regions (VRs).

\subsection{Background estimation in the zero-lepton signal regions}
The main SM background in the b0L signal regions is from the production of \Zjets followed by invisible decays of the $Z$ boson. The production of top quark pairs, single top quarks and \Wjets also results in important backgrounds, with their relative contributions depending on the specific SR considered. Full details of the CR definitions are given in Tables~\ref{tab:CR0Lab} and \ref{tab:CR0Lc}.

\begin{table}[!htpb]
  \centering

  \caption{Summary of the event selection in each control region corresponding to \bZerLSRA\ and \bZerLSRB.
The term lepton is used in the table to refer to signal electrons and muons. Jets ($j_1$, $j_2$, $j_3$ and $j_4$) and leptons ($\ell_1$ and $\ell_2$) are labelled with an index corresponding to their decreasing order in \pt.}
\vspace{0.2cm}
  \label{tab:CR0Lab}
\begin{small}

      \begin{tabular*}{\textwidth}{@{\extracolsep{\fill}}lc@{}  c c c | c c c}
    \hline
    \toprule
    b0L-      & CRzA &   CRttA &  CRstA & CRwA & CRzB & CRttB & CRwB \\
    \midrule
    Number of leptons ($\ell=e,\ \mu$) & 2 SFOS & 1 & 1 & 1 & 2 SFOS & 1 & 1 \\
    $\pt(\ell_1)$ [\GeV] & $>90$ & $>27$ & $>27$ & $>27$ & $>27$ & $>27$ & $>27$ \\
    $\pt(\ell_2)$ [\GeV] & $>20$ & -- & -- & -- & $>20$ & -- & -- \\
    $\mll$ [\GeV] & [76 --106] & -- & -- & -- & [76--106] & -- & -- \\ 
    \Njets(\pt$>35$~\GeV) & 2--4 & 2--4 & 2--4 & 2--4 & 2--4 & 2--4 & 2--4 \\
    $\pt(j_1)$ [\GeV]       & $>50$ & $>130$ & --&  $>130$ & $>50$ & $>50$ & $>50$ \\
    $\pt(j_2)$ [\GeV]        & $>50$ & $>50$ & $>50$ & $>50$ & $>50$ & $>50$ & $>50$ \\
    $\pt(j_4)$ [\GeV]          &  $<50$ &    $<50$ &  $<50$ &  $<50$ & -- & -- & -- \\
    \bjets & $j_1$ and $j_2$ & $j_1$ and $j_2$ & $j_1$ and $j_2$ & $j_1$ & any 2 & any 2 & any 2 \\
    \etmiss\   [\GeV]         &  $<100$ & $>200$ & $>200$ & $>200$ & $<100$ & $>100$ & $>100$ \\
    \etmisscor\ [\GeV]  & $>100$ & -- & -- & -- & $>200$ & -- & -- \\
    \met/\meff         &  $>0.25$ & $>0.25$  & $>0.25$ & $>0.25$ & -- & -- & -- \\
    \dphiminfour       &  --      & $>0.4$   & $>0.4$  & $>0.4$  & $>0.4$  & $>0.4$  & $>0.4$  \\
    $\mt$ [\GeV] & -- & -- & -- & $>30$ & -- & $>30$ & $>30$ \\
    $\mbb$ [\GeV] & $>200$ & $<200$ & $>200$ & $\mbj>200$ & -- & -- & -- \\
    $\mct$ [\GeV] & $>250$ & $>250$ & $>250$ & $>250$ & -- & -- & -- \\
    $\mlbmin$ [\GeV] & -- & -- & $>170$ & -- & -- & -- & -- \\
    $\minmtfour$ [\GeV] & -- & -- & -- & -- & $>200$ & $>200$ & $>250$ \\
    $\Delta\phi(b_1,\met)$ &    -- & -- & -- & -- &  -- & $<2.0$ & $<2.0$ \\ 
    $\Delta\phi(b_2,\met)$ &    -- & -- & -- & -- &  -- & $<2.5$ & -- \\ 
    \hline
  \end{tabular*}
  \end{small}
\end{table}

\begin{table}[!htpb]
  \centering

  \caption{Summary of the event selection in each control region corresponding to \bZerLSRC.
The term lepton is used in the table to refer to signal electrons and muons. Jets ($j_1$, $j_2$, $j_3$ and $j_4$) and leptons ($\ell_1$ and $\ell_2$) 
are labelled with an index corresponding to their decreasing order in \pt.}
\vspace{0.2cm}
  \label{tab:CR0Lc}
\begin{small}
\begin{tabular*}{\textwidth}{@{\extracolsep{\fill}}lc@{} c c c}
    \hline
    \toprule
    b0L-      & CRzC &   CRttC &  CRwC \\
    \midrule
    Number of leptons ($\ell=e,\ \mu$) & 2 SFOS & 1 & 1 \\
    $\pt(\ell_1)$ [\GeV] & $>27$ & $>27$ & $>27$ \\
    $\pt(\ell_2)$ [\GeV] & $>20$ & -- & -- \\
    $\mll$ [\GeV] & [76 --106] & -- & -- \\ 
    \Njets(\pt$>20$~\GeV) & 2--5 & 2--5 & 2--5 \\
    Leading jet \pt [\GeV]       & $> 250$ & $> 500$ & $> 500$ \\
    \bjets  & $j_2$ and ($j_3$ or $j_4$) & $j_2$ and ($j_3$ or $j_4$) & $j_2$ \\ 
    \etmiss\   [\GeV]         &  $<100$ & $>100$ & $>100$ \\ 
    \etmisscor\ [\GeV]  & $>200$ & -- & -- \\
    $\mt$ [\GeV] & -- & $>30$ & $[30$--$120]$ \\ 
    $\meff$ [\GeV] & $>500$ & $>1300$ & $>500$ \\ 
    $\mjj$ [\GeV] & $>200$ & $>200$ & $>200$ \\
    $\htfour$ [\GeV] & $<70$ & $<70$ & $<70$ \\
    ${\cal A}$ & $>0.5$ & $>0.5$ & $>0.8$ \\
    $\Delta\phi(j_1,\met)$ &   $>2.5$ & $>2.5$ & $>2.5$ \\
    \hline
  \end{tabular*}
  \end{small}
\end{table}

Three same-flavour opposite-sign (SFOS) two-lepton (electron or muon) control regions with dilepton invariant mass near the $Z$ boson mass ($76 < \mll < 106$~\GeV) and two \btagged\ jets provide data samples dominated by Z boson production. Signal leptons are considered, with the threshold for the second lepton \pt\ loosened to 20~\GeV. For these control regions, labelled in the following as \bZerLCRAZ, \bZerLCRBZ\ and
\bZerLCRCZ, the \pt of the leptons is added vectorially to the \ptmiss to mimic the expected missing transverse momentum spectrum of $Z \rightarrow \nu \bar{\nu}$ events, and is indicated in the following as \etmisscor (lepton corrected). In addition, a selection is applied to the uncorrected \etmiss of the event, in order to further enhance the $Z$ boson contribution.

Events with one charged lepton in the final state are used to define control regions dominated by \Wjets\ and top quark production by requiring either one or two \btagged\ jets, respectively. Selections on the variable $\mt$ are used to ensure that the lepton originates from a $W$ decay. For the CRs corresponding to \bZerLSRA,  the contribution from $\ttbar$ and single top quark production are separated by applying the selection $\mbb<200$~\GeV\ and $\mbb>200$~\GeV, respectively. To further enhance the single-top-quark contribution, a selection on the minimum invariant mass of the lepton and one of the \bjets, $\mlbmin>170$~\GeV\ is applied. For the CRs corresponding to \bZerLSRB, selections on the azimuthal angle between the \bjets\ and the \etmiss\ value are applied to enhance the $\ttbar$ and \Wjets\ contributions, while the single-top-quark background is estimated from MC simulation. The CRs corresponding to the \bZerLSRC\ are defined with one or two \bjets\ to enhance the $\ttbar$ and \Wjets\ contributions, respectively. Finally, the single top quark production is estimated using the MC normalization.

The contributions from dibosons ($WW, WZ, ZZ$), $t\bar{t}$ production associated with $W$ and $Z$ bosons, and other rare backgrounds are estimated 
from MC simulation for both the signal and the control regions and included in the fit procedure, and are allowed to vary within 
their normalization uncertainty. The background from multijet production is estimated from data using a procedure described in detail 
in \Refe~\cite{SUSY-2011-20} and modified to account for the heavy flavour of the jets. 
The contribution from multijet production in all regions is found to be negligible.

In total, four CRs are defined for the \bZerLSRA\ to estimate the contributions from \Wjets, \Zjets, $\ttbar$ and single top quark production independently, while three CRs are defined for each of the \bZerLSRB\ and \bZerLSRC\ to estimate  \Wjets, \Zjets and $\ttbar$. The $\etmiss$ distribution in b0L-CRwA and b0L-CRzC is shown in Figures~\ref{fig:CRplots_W} and \ref{fig:CRplots_Z}, where good agreement with the SM prediction is achieved after the background-only fit. 
The yields in all these CRs are shown in Figure~\ref{fig:CRpulls} and compared to the MC predictions before the likelihood fit is performed, including only the statistical uncertainty of the MC samples.    
The bottom panel shows the value of the normalization factors, $\mu$, used for each of the backgrounds fitted and given taking into account statistical and detector-related systematic uncertainties. 

As a further validation, two alternative methods are used to estimate the \Zjets\ contribution. The first method exploits the similarity of the \Zjets\ and \gamjets\ processes~\cite{SUSY-2011-20}. For a photon with \pt~significantly larger than the mass of the $Z$ boson, the kinematics of \gamjets\ events strongly resemble those of \Zjets\ events. A set of dedicated control regions is defined by requiring one isolated photon with $\pt>145$~\GeV.  The \pt of the photon is vectorially added to the \ptmiss, and the magnitude of this sum is used to replace the \etmiss-based selections. The yields are then propagated to the SRs using a reweighting factor derived using the MC simulation. This factor takes into account the different kinematics of the two processes and residual effects arising from the different geometrical acceptance and reconstruction efficiency for photons. In the second alternative method, applied to \bZerLSRA\ only, the MC simulation is used to verify that the shape of the $\mct$ distribution for events with no \btagged\ jets is compatible with the shape of the $\mct$ distribution for events where two \btagged\ jets are present. A new highly populated \Zjets\ CR is defined, selecting $Z \rightarrow \ell \ell$ events with no \btagged\ jets. The $\mct$ distribution in this CR is constructed using the two leading jets and is used to estimate the shape of the $\mct$ distribution in the \bZerLSRA, whilst the normalization in SRA is rescaled based on the ratio in data of $Z \rightarrow \ell \ell$ events with no \btagged\ jets to events with two \btagged\ jets. Additional MC-based corrections are applied to take into account the two-lepton selection in this CR.
The two alternative methods are in agreement within uncertainties with the estimates obtained with the profile likelihood fit to the control regions. Experimental and theoretical systematic uncertainties in the estimates from the nominal and alternative methods are taken into account (see Section~\ref{sec:systematics}). 

\subsection{Background estimation in the one-lepton signal regions}
The main SM background in the b1L signal regions is the production of $\ttbar$ and single-top-quark events in the $Wt$ channel. Two control regions (\bOneLCRAtt\ and \bOneLCRBtt) where the $\ttbar$  production is enhanced are defined by inverting the $\amtTwo$ selection. In the case of \bOneLCRAtt\ the $\mbb$ selection is also inverted, while for \bOneLCRBtt\ the $\mtbmin$ requirement is inverted. To allow a statistical combination of the results from the \bZerLSRA\ and \bOneLSRA\ regions the corresponding $\ttbar$ CRs are defined to be orthogonal via the $\mct$ selection. The single-top-quark contribution is estimated with the same CR employed by the b0L analysis.  In the case of  \bOneLSRB\ the production of \Wjets\ is no longer negligible, and is estimated by using a dedicated control region \bOneLCRBW, where only one \btagged\ jet is required. In total, two CRs are used to estimate the event yields in \bOneLSRA\ and three CRs to estimate the yields in \bOneLSRB. Full details of the CR selections are given in Table~\ref{tab:CR1L}.
The  distribution of $\mbb$ in b1L-CRstA and of $\mt$ in b1L-CRttB are presented in Figures~\ref{fig:CRplots_st} and \ref{fig:CRplots_tt} to show the level of agreement achieved after the background-only fit. \\

The yields in all these CRs are also shown in Figure~\ref{fig:CRpulls} and compared to the direct MC prediction before the likelihood fit is performed. 
The normalization parameters reported for each SR and SM background process include the statistical and 
detector-related systematic uncertainties.  
The decrease of the $\mu^{}_{\ttbar}$ parameter from SRA to SRC is related to mismodelling 
in the description of $\ttbar$ processes by \powheg+\pythia6 MC samples. Previous analyses~\cite{Aad:2015mia} 
also found normalization factors considerably smaller than unity for $\ttbar$ background processes in 
similar regions of phase space. 
 The $W$+jets and $Z$+jets normalization factors are larger than unity. This is possibly related to the fact that in the default \sherpa~2.2.1 
the heavy-flavour production fractions are not consistent with the measured values~\cite{ATL-PHYS-PUB-2017-007}.

\begin{table}[!htpb]
  \centering

  \caption{Summary of the event selection in each control region corresponding to the b1L signal regions.
The term lepton is used in the table to refer to signal electrons and muons. Jets ($j_1$, $j_2$, $j_3$ and $j_4$)
are labelled with an index corresponding to their decreasing order in \pt.}
\vspace{0.2cm}
  \label{tab:CR1L}
\begin{small}
\begin{tabular*}{\textwidth}{@{\extracolsep{\fill}}lc@{} c | c c c }
    \hline
    \toprule
    b1L-      & CRttA &   CRstA &  CRttB & CRstB & CRwB  \\
    \midrule
    Number of leptons ($\ell=e,\ \mu$) &  1 & 1 & 1 & 1 & 1 \\
    $\pt(\ell)$ [\GeV] & $>27$ &  $>27$ &  $>27$ &  $>27$ &  $>27$ \\
    \Njets(\pt$>35$~\GeV) & $\geq 2$ & $[2$--$4]$ & $\geq 2$ & $\geq 2$ & $\geq 2$ \\
    $\pt(j_1)$ [\GeV]       & $> 35$ & $> 130$ & $> 35$ & $> 35$ & $> 35$   \\
    $\pt(j_2)$ [\GeV]       & $> 35$ & $> 50$ & $> 35$ & $> 35$ & $> 35$   \\
    $\pt(j_4)$ [\GeV]       & $> 35$ & $ [35$--$50]$ & -- & -- & --   \\
    \dphiminfour      & $>0.4$ & $>0.4$  & $>0.4$  & $>0.4$  & $>0.4$ \\
    \bjets & any 2 & $j_1$ and ($j_2$ or $j_3$ or $j_4$) & any 2 & any 2 & any 1 \\
    $\mbb$ [\GeV] & $<200$ & $>200$ & $<200$ & $>200$ & $>200$  \\
    $\mlbmin$ [\GeV] & $<170$ & $>170$ & $<170$ & $>170$ & $<170$  \\
    \etmiss\   [\GeV]         &  $>200$ & $>200$ & $>200$ & $>200$ & $>200$ \\ 
    $\mt$ [\GeV] & $>140$ & -- & $>120$ &  $[30$--$120]$ & $[30$--$120]$ \\ 
    $\amtTwo$ [\GeV] & $<250$ & -- & $<200$ & -- & $>200$ \\ 
    $\meff$ [\GeV] & $>300$ & -- & -- & -- & -- \\
    $\mct$ [\GeV] & $<250$ & $>250$ & -- & -- & -- \\
    $\etmiss / \sqrt{\HT}$ [\GeV$^{1/2}$] & $>8$ & -- & $>8$ & $>8$ & $>8$ \\
    $\etmiss / \meff$ & -- & $>0.25$ & -- & -- & -- \\
    $\minmtb$ [\GeV] & -- & -- & $<200$ & $>200$ & $>200$ \\
    $\Delta\phi(b_1,\met)$ &  -- & -- & $>2.0$ & $>2.0$ & $>2.0$ \\
    \hline
  \end{tabular*}
  \end{small}
\end{table}

\begin{figure}[t]
\begin{center}
\subfloat[\label{fig:CRplots_W}]{
\includegraphics[width=0.49\textwidth]{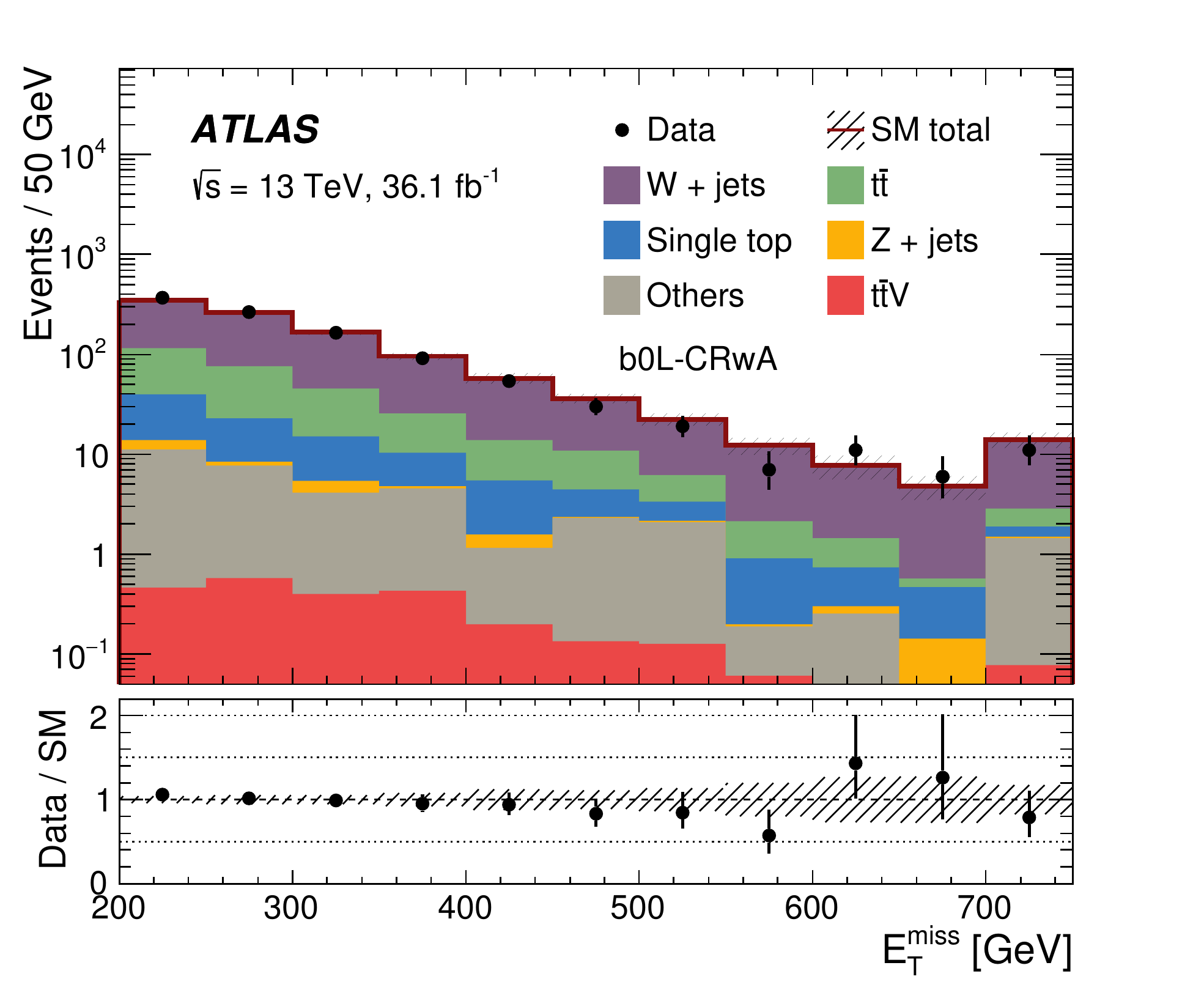}
}
\subfloat[\label{fig:CRplots_Z}]{
\includegraphics[width=0.49\textwidth]{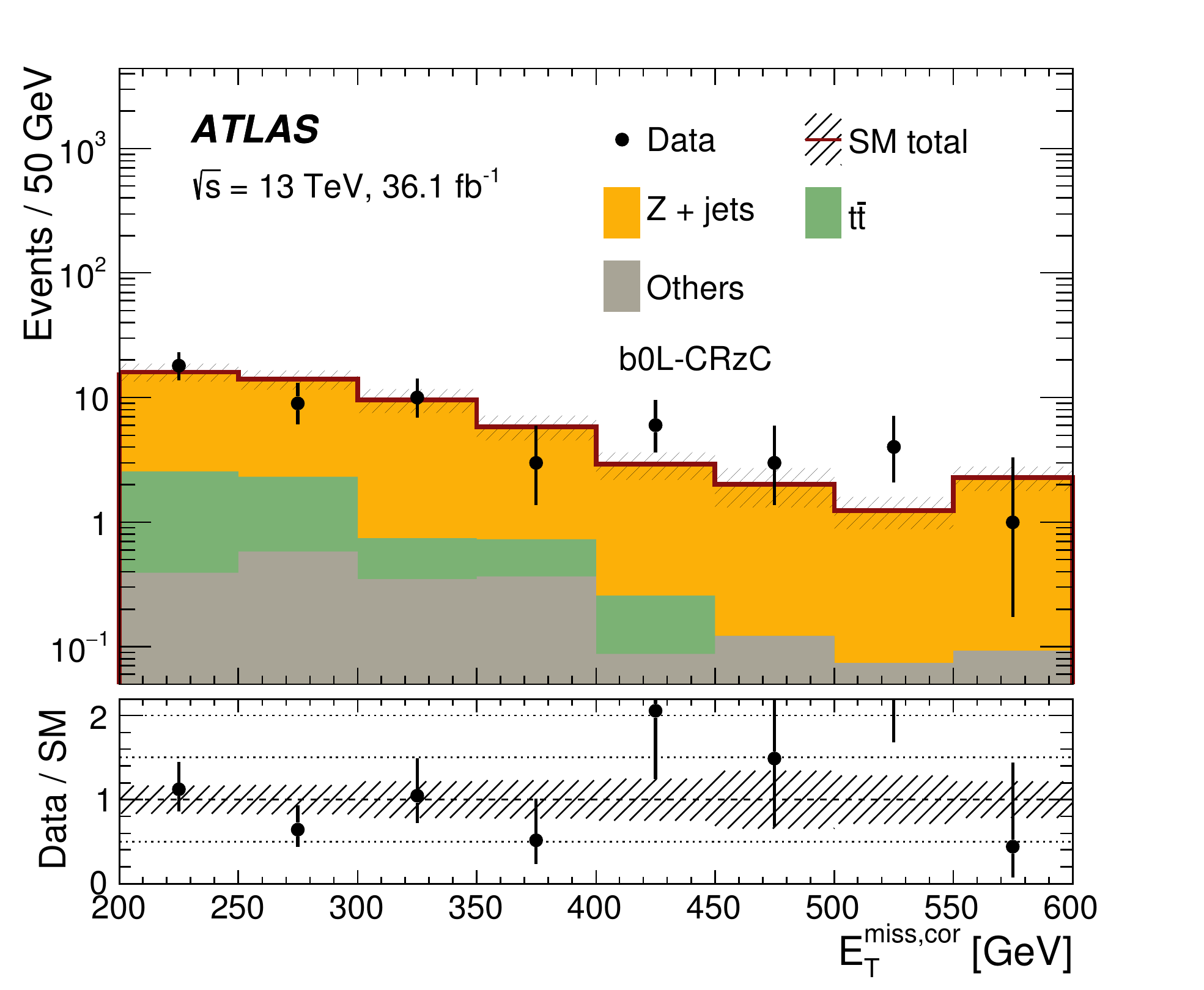}
} \\
\subfloat[\label{fig:CRplots_st}]{
\includegraphics[width=0.49\textwidth]{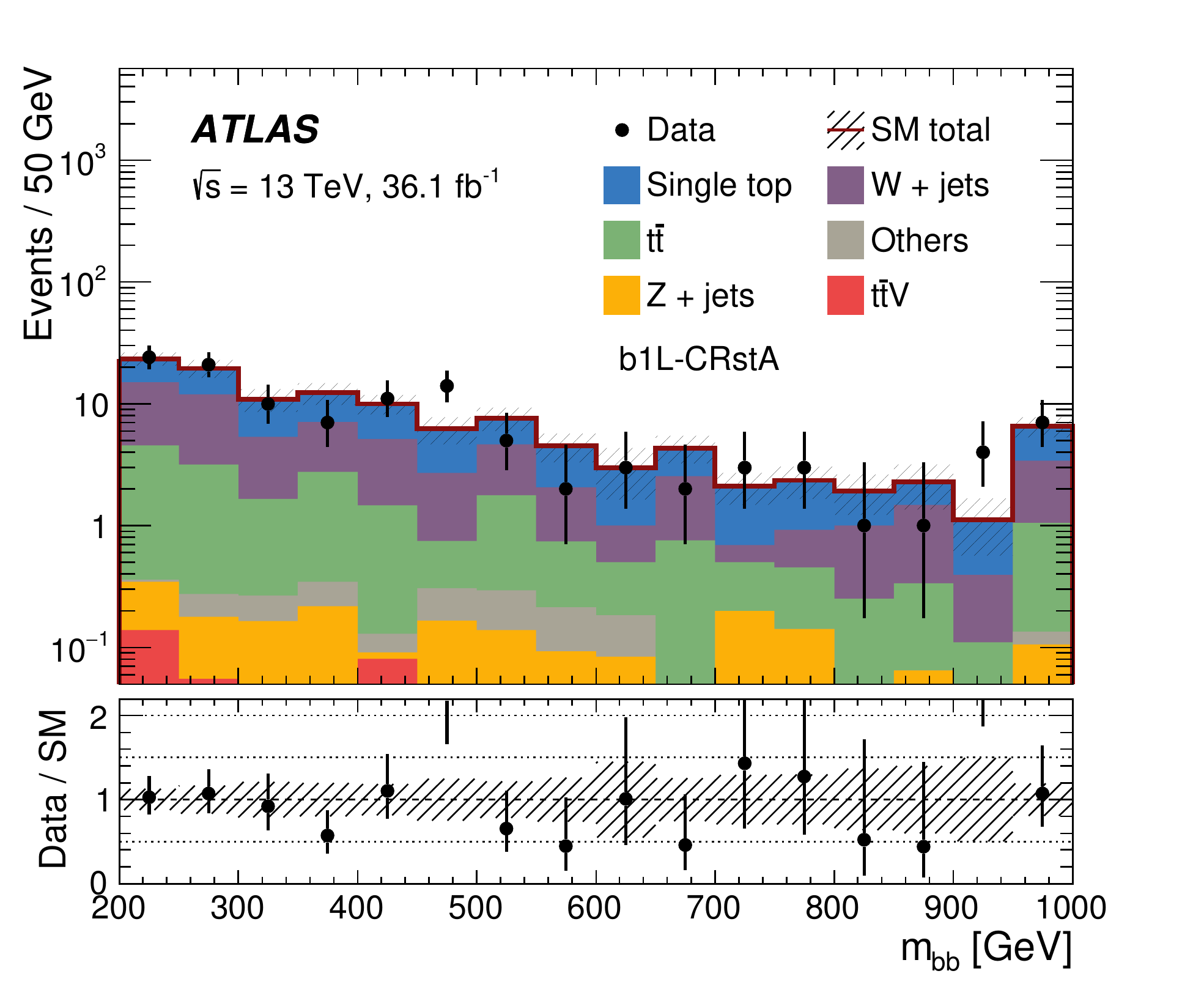}
}
\subfloat[\label{fig:CRplots_tt}]{
\includegraphics[width=0.49\textwidth]{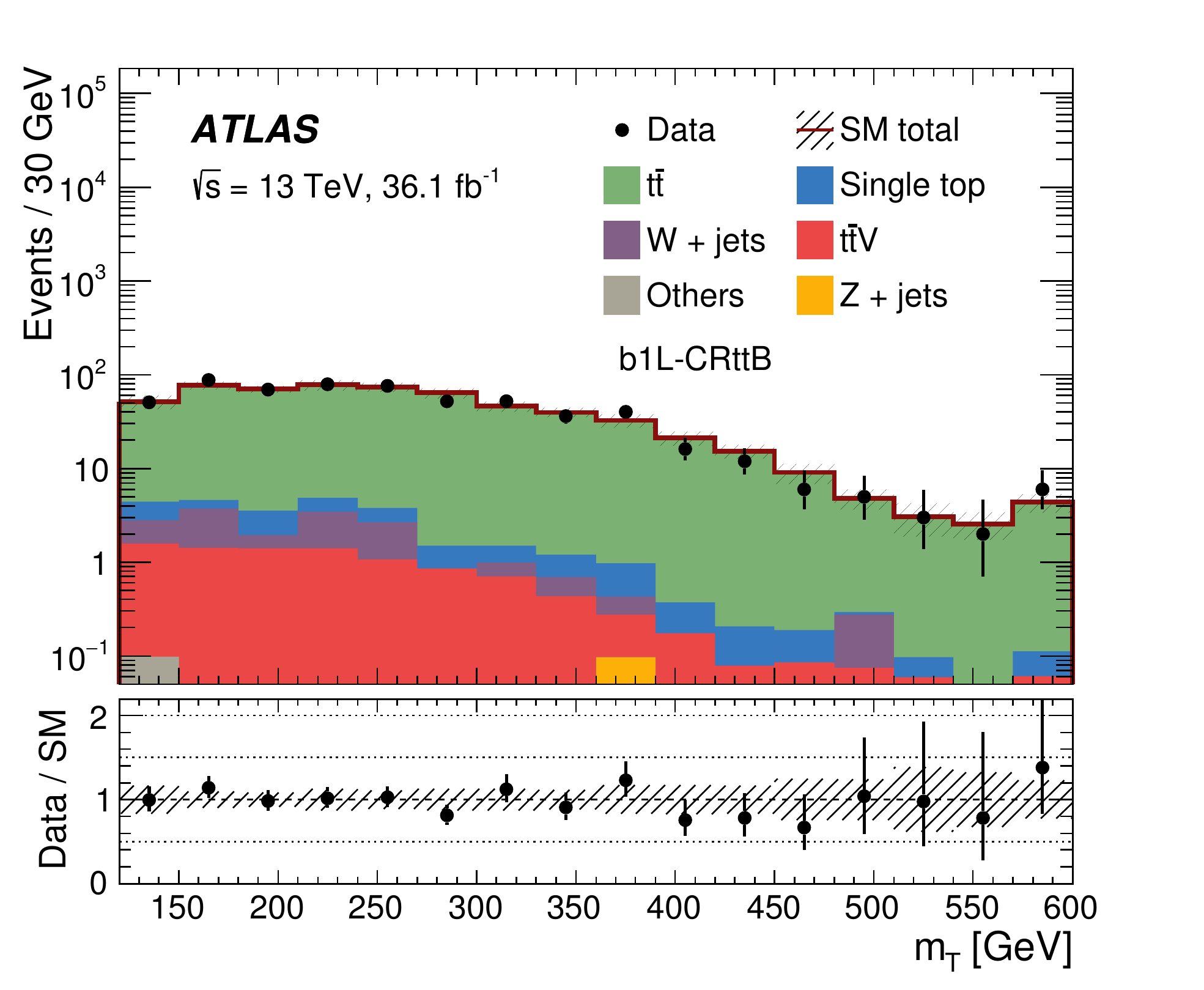}
}
\end{center}
\caption{\label{fig:CRplots} Example kinematic distributions in some of the control regions. 
(a) $\etmiss$ in b0L-CRwA, 
(b) $\etmisscor$ in b0L-CRzC, 
(c) $\mbb$ in b1L-CRstA and
(d) $\mt$ in b1L-CRttB. In all distributions the MC normalization is rescaled using the results from the background-only fit, showing good agreement between data and the predicted SM shapes. The contributions from diboson, multijet and rare backgrounds are collectively called ``Others''. 
The shaded-grey band shows the detector-related systematic uncertainties and the statistical uncertainties of the MC samples as detailed in Section~\ref{sec:systematics} and the last bin includes overflow events.
}
\end{figure}

\begin{figure}[t]
\begin{center}
\includegraphics[width=0.99\textwidth]{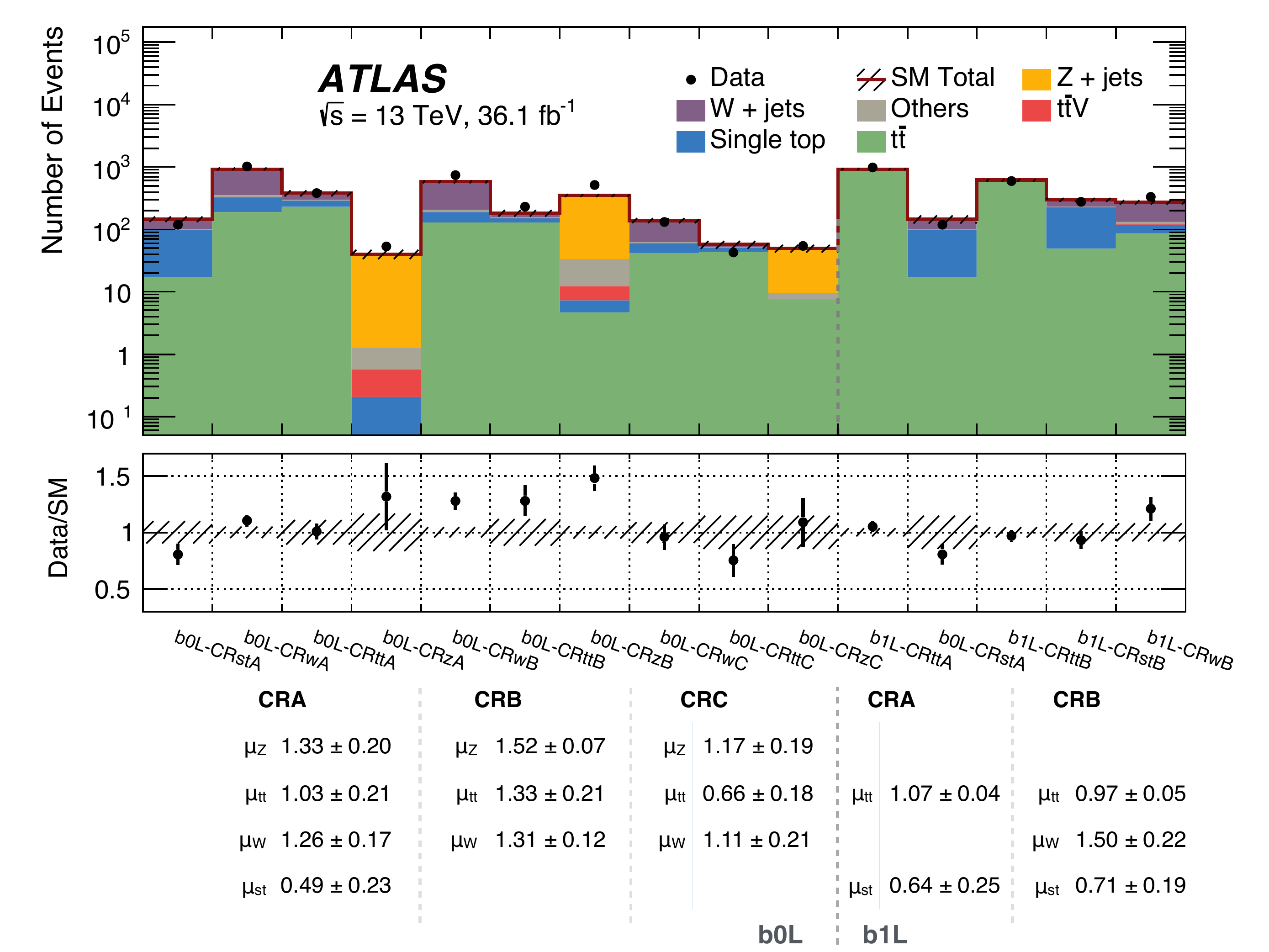}
\end{center}
\caption{ Results of the likelihood fit to the CRs associated with each of the b0L and b1L SRs. In the top panel the normalization of 
 the backgrounds is obtained from MC simulation and is the input value to the fit. 
The contributions from diboson, multijet and rare backgrounds are collectively called ``Others''. 
The panels at the bottom show  the ratio of the observed events in each CR to the MC estimate, and the value of the normalization factors ($\mu$) obtained for each of the backgrounds fitted. 
The uncertainty band around the MC prediction includes only the statistical uncertainty of the MC samples. The normalization factors $\mu$ are presented for each region and SM background process and take into account statistical and detector-related systematic uncertainties. }
\label{fig:CRpulls}
\end{figure}

\subsection{Validation regions}
The results of the background-only fit to the CRs are extrapolated to a set of VRs defined to be similar to the SRs, with some of the selection criteria modified to enhance the background contribution, while maintaining a small signal contribution. For each SR, one or more VRs are defined starting from the SR definition and inverting or changing some of the selections as summarized in Table~\ref{tab:VR}. 

\begin{table}[!htpb]
  \centering

  \caption{Summary of the VRs used in the analysis. Each VR (left column) corresponds to a SR (middle column) defined in Tables~\ref{tab:SR0l} and \ref{tab:SR1L},  with a few selection requirements changed (right column) 
to ensure the selection has low efficiency for the expected signal.
}
\vspace{0.2cm}
  \label{tab:VR}
\begin{small}
\begin{tabular*}{\textwidth}{@{\extracolsep{\fill}}lc@{} c }
    \hline
    \toprule
    VR      & Corresponding SR &   Selection changes  \\
    \midrule
    b0L-VRmctA & b0L-SRA & $\minmtfour<250$~\GeV, $150<\mct<250$~\GeV \vspace{0.15cm} \\
    b0L-VRmbbA & b0L-SRA & $\minmtfour<250$~\GeV, $100<\mbb<200$~\GeV \vspace{0.15cm} \\
    b0L-VRzB &   b0L-SRB & $\mct<250$~\GeV, $200<\minmtfour<250$~\GeV, \\
             &           & ${\cal A}<0.8$, no selection on $\dphibone$ and $\dphibtwo$ \vspace{0.15cm} \\
    b0L-VRttB &  b0L-SRB & $\mct<250$~\GeV, $150<\minmtfour<200$~\GeV, ${\cal A}<0.8$ \vspace{0.15cm} \\
    b0L-VRttC & b0L-SRC  & $\mct<250$~\GeV, $\minmtfour<250$~\GeV, $0.6<{\cal A}<0.8$ \vspace{0.15cm} \\
    \hline
    b1L-VRamt2A & b1L-SRA300-2j & $30<\mt<140$~\GeV, $\mbb<200$~\GeV \vspace{0.15cm}\\
    b1L-VRmbbA  & b1L-SRA300-2j & $\amtTwo<250$~\GeV \vspace{0.15cm}\\
    b1L-VRamt2B & b1L-SRB       & $\dphibone>2.0$,  $\mbb>200$~\GeV \vspace{0.15cm}\\
    b1L-VRmbbB  & b1L-SRB       & $\dphibone>2.0$, $30<\mt<120$~\GeV \\
  \hline
  \end{tabular*}
  \end{small}
\end{table}

The number of events predicted by the background-only fit is compared to the data in the upper panel of Figure~\ref{fig:VRpulls}.  
The pull, defined by the difference between the observed number of events ($n_{\textrm{obs}}$) and the predicted background yield ($n_{\textrm{pred}}$) divided by the total uncertainty ($\sigma_{\textrm{tot}}$), is shown for each region in the lower panel. No evidence of significant background mismodelling is observed in the VRs.

\begin{figure}[t]
\begin{center}
\includegraphics[width=0.99\textwidth]{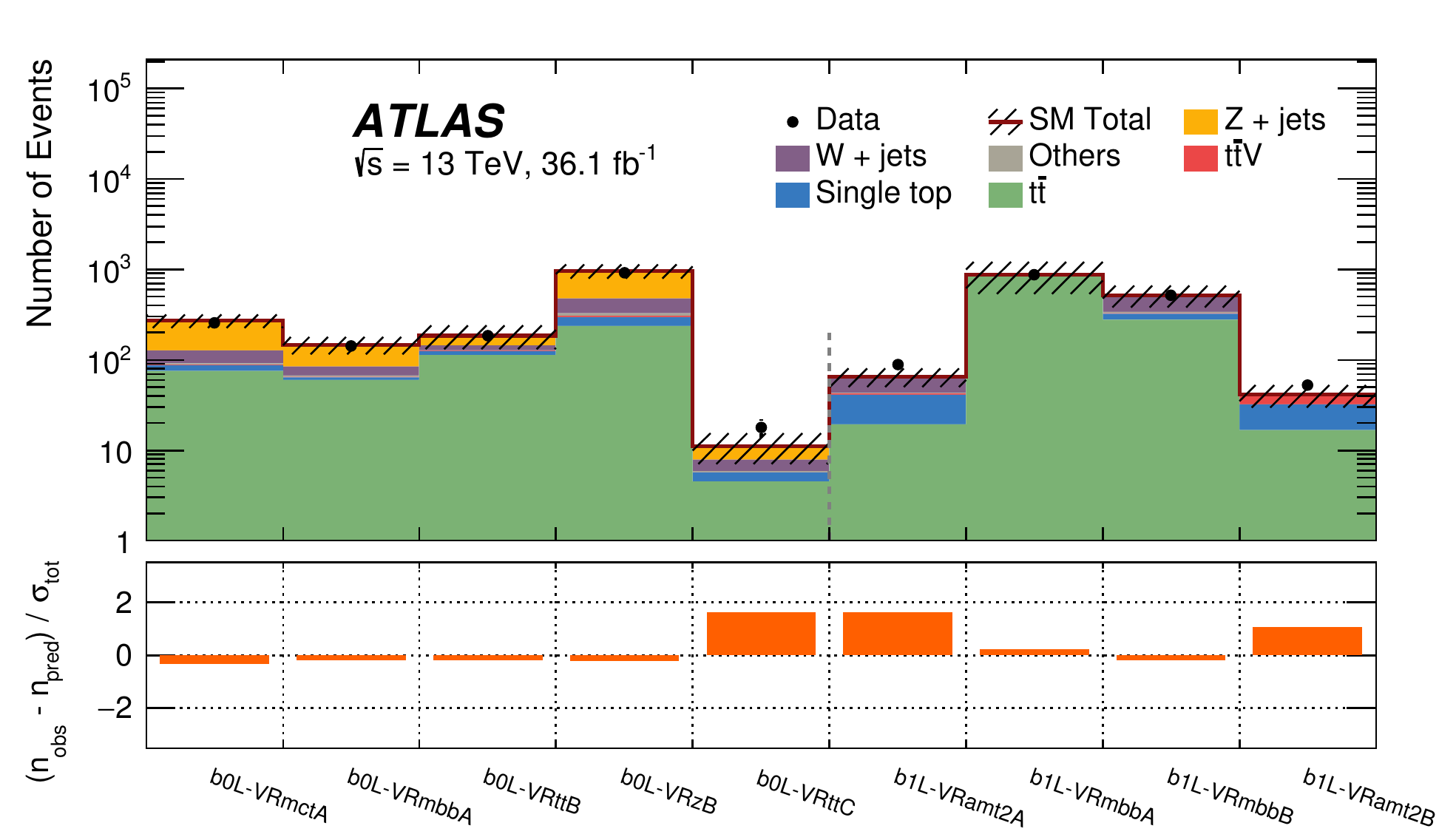}
\end{center}
\caption{ Results of the likelihood fit extrapolated to the VRs associated with the b0L and b1L analyses. The normalization of the backgrounds is obtained from the fit to the CRs. The upper panel shows the observed number of events and the predicted background yield. 
The contributions from diboson, multijet and rare backgrounds are collectively called ``Others''. 
All uncertainties defined in Section~\ref{sec:systematics} are included in the uncertainty band.  The lower panel shows the pulls in each VR, where
$\sigma_{\textrm{tot}}$ is the error on the background estimation as a sum in quadrature of the systematic uncertainty and the statistical uncertainty on the estimate. }
\label{fig:VRpulls}
\end{figure}

\section{Systematic uncertainties}
\label{sec:systematics}
Several sources of experimental and theoretical systematic uncertainty in the signal and background 
estimates are considered in these analyses. 
Their impact is reduced through the normalization of the dominant backgrounds in the control regions defined with 
kinematic selections resembling those of the corresponding signal region (see \Sect~\ref{sec:background}). 
Experimental and theoretical uncertainties are included as nuisance parameters with Gaussian constraints 
in the likelihood fits, taking into account correlations between different regions. 
Uncertainties due to the numbers of events in the CRs are also introduced in the fit for each region. 
The dominant contributions are summarized in Table~\ref{tab:syst_summary}.

\begin{table}[htbp]
\begin{center}
\setlength{\tabcolsep}{0.0pc}
\caption{Summary of the dominant experimental and theoretical uncertainties for each signal region in zero-lepton and one-lepton channels. 
Uncertainties are quoted as relative to the total SM background predictions, with a range indicated for the three b0L-SRAs and the two 
b1L-SRAs. 
For theoretical modelling, uncertainties per dominant SM background process are quoted. 
The individual uncertainties can be correlated, and do not necessarily add in quadrature to the total background uncertainty.}
\label{tab:syst_summary}
{\small
  \begin{tabular*}{\textwidth}{@{\extracolsep{\fill}}l ccc ccc}
    \noalign{\smallskip}\hline\noalign{\smallskip}
  \textbf{Source \textbackslash~Region}    &     b0L-SRAx    & b0L-SRB & b0L-SRC &  b1L-SRAx & b1L-SRB & b1L-SRA300-2j    \\[-0.05cm]
     \noalign{\smallskip}\hline\noalign{\smallskip}
  Experimental uncertainty  &  &  & & & & \\ 
  \midrule
  JES  & 2.3\phantom{0}--\phantom{0}3.4\% & 5.7\% & 4.3\% & \phantom{1}1.2\phantom{0}--\phantom{0}1.5\% & 0.9\% & 6.9\% \\
  JER  & 0.9\phantom{0}--\phantom{0}3.3\% & 3.5\% & 11\% & \phantom{1}5.3\phantom{0}--\phantom{0}8.6\% & 0.9\% & 4\% \\ 
  $b$-tagging & 3.3\phantom{0}--\phantom{0}4.3\% & 7.5\% & 4.7\% & \phantom{1}6.1\phantom{0}--\phantom{0}6.3\% & 2\% & 6.6\% \\ 
  \midrule
  Theoretical modelling uncertainty  &  & & & & &  \\ 
  \midrule 
  $Z$+jets & \phantom{0}9.6\phantom{0}--\phantom{0}12\%\phantom{.3} & 13\% & 11\% & - & - & - \\ 
  $W$+jets & 3.4\phantom{0}--\phantom{0}5.2\% & 4.7\% & 7.6\% & \phantom{1}1.3\phantom{0}--\phantom{0}1.6\% & 8.6\% & 7.9\% \\ 
  Top production & 2.2\phantom{0}--\phantom{0}3.1\% & 6\% & 3.6\% & 19\% & 13\% & 22\% \\ 
    \noalign{\smallskip}\hline\noalign{\smallskip}
    \end{tabular*}
     }
     \end{center}
     \vspace{-0.2cm}
     \end{table}
The dominant detector-related systematic effects are due to the uncertainties in the jet energy scale (JES)~\cite{PERF-2016-04}  
and resolution (JER)~\cite{PERF-2011-04}, and in the $b$-tagging efficiency and mis-tagging rates. The latter  
are estimated by varying the $\eta$-, \pt- and flavour-dependent scale factors applied to each jet in the simulation within 
a range that reflects the systematic uncertainty in the measured tagging efficiency and mis-tag rates in 13~\TeV\ data. 
 The uncertainties associated with lepton and photon reconstruction and energy measurements are 
also considered but have a negligible impact on the final results. Lepton, photon and jet-related uncertainties are propagated 
to the \etmiss\ calculation, and additional uncertainties are included in the energy scale and resolution of the soft term. 
 
Uncertainties in the modelling of the SM background processes from MC simulation and their theoretical cross-section uncertainties 
are also taken into account. The dominant uncertainty arises from $Z$+jets MC modelling for b0L-SRs and $\ttbar$ and single-top 
modelling (collectively referred to as ``Top production'' in Table~\ref{tab:syst_summary}) for b1L-SRs.  
The \Zjets\ (as well as \Wjets) modelling uncertainties are estimated by considering different merging (CKKW-L) 
and resummation scales using alternative samples, PDF variations from the NNPDF30NNLO replicas~\cite{Gleisberg:2008ta},  
as well as an envelope formed from seven-point scale variations of the renormalization and factorization scales. 
The various components are added in quadrature.
A 40\% uncertainty~\cite{STDM-2011-21} is assigned to the heavy-flavour jet content in $W$+jets background, which is estimated from MC simulation 
in the one-lepton channel control regions. 
For b0L-SRA, b0L-SRC and b1L-SRB the uncertainty accounts for the different requirements on \bjets between the signal regions 
and the corresponding control regions.   

Theoretical and modelling uncertainties of the top quark pair and single-top-quark ($Wt$) backgrounds are computed as the difference between the prediction from nominal samples and those from 
additional samples differing in generator or parameter settings. Hadronization and PS uncertainties are estimated using samples 
generated using \powhegbox v2 and showered by \herwigpp v2.7.1~\cite{Bahr:2008pv} with the UEEE5~\cite{Gieseke:2012ft}  underlying-event tune. 
Uncertainties related to initial- and final-state radiation modelling, PS tune and (for \ttbar~only) choice of $h^{}_{\mathrm{damp}}$ parameter 
in \powhegbox v2 are estimated using alternative settings of the event generators. Finally, an alternative generator \madgraph with showering 
by \herwigpp v2.7.1 is used to estimate the event generator uncertainties. 
One additional uncertainty stems from the modelling of the interference between the \ttbar\ and $Wt$ processes at NLO. 
Predictions from an inclusive $WWbb$ sample generated at LO using \madgraph\ are compared with the sum of 
the \ttbar\ and $Wt$ predictions, and differences from the nominal predictions are taken as systematic uncertainties.  

Uncertainties in backgrounds such as diboson and $ttV$ are also estimated by comparisons of the nominal sample 
with alternative samples differing in generator or parameter settings (\powheg v2 with showering by \pythia~v8.210 for dibosons; 
renormalization and factorization scale and A14 tune variations for $ttV$) and contribute less than 5\% to the total uncertainty.  
The cross-sections used to normalize the MC yields to the highest order available are varied according to the scale uncertainty 
of the theoretical calculation. The cross-section uncertainties are 5\% for $W$ boson, $Z$ boson and top quark pair production, 
6\% for dibosons, and 13\% and 12\% for $ttW$ and $ttZ$, respectively.  
Finally, a conservative 100\% systematic uncertainty associated to the multijet background estimate is considered and found 
to have a negligible effect. 
 

For the SUSY signal processes, both the experimental and theoretical uncertainties in the expected signal yield are considered. 
Experimental uncertainties are found to be between 15\% and 30\% across the $\bone$--$\neut$ mass plane for exclusive 
$\bone \rightarrow b\neut$ decays and between 10\% and 25\% for models where bottom squarks decay with a significant branching ratio 
as $\bone \rightarrow t\chipm$, assuming the one-lepton channel selection. In all SRs, they are largely dominated by the 
uncertainty in the $b$-tagging efficiency. Theoretical uncertainties in the NLO+NLL cross-section are calculated for each 
SUSY signal scenario and are dominated by the uncertainties in the renormalization and factorization scales, followed 
by the uncertainty in the PDF. They vary between 15\% and 25\% for bottom squark masses in the range between 400~\GeV\ and 1100~\GeV.  
Additional uncertainties in the acceptance and efficiency due to the modelling of initial-state radiation and scale variations 
in SUSY signal MC samples are also taken into account and contribute up to about 10\%.

\section{Results and interpretation}
\label{sec:results}
Tables~\ref{tab:fit_0LSR} and ~\ref{tab:fit_1LSR} report the observed number of events and the SM prediction
after the background-only fit for each signal region in the zero-lepton and one-lepton channels, respectively. 
The background-only fit results are compared to the pre-fit predictions based on MC simulation. 
The largest background contribution in b0L-SRs arises from $Z\rightarrow \nu\bar{\nu}$  produced in 
association with $b$-quarks followed by $W$+jets production, whilst top quark and $W$+jets 
production dominates SM predictions for b1L-SRs.  
The results are also summarized in Figure~\ref{fig:SRpulls}, where the pulls for each of the SRs are also presented. 
No significant excess above the expected Standard Model background yield is observed, although 
b1L-SRA300-2j presents a discrepancy between data and SM predictions of about 1.5$\sigma$.  

\begin{table}
\begin{center}
\setlength{\tabcolsep}{0.0pc}
{\small
\caption{
Fit results in the b0L signal regions. The background normalization parameters are obtained 
 from the fit in the control regions and are applied to the SRs. 
Smaller backgrounds such as diboson, $ttV$, multijet and rare processes are indicated as ``Others''.   
The individual 
uncertainties, including statistical, detector-related and theoretical systematic components, are symmetrized and can be correlated. 
They do not necessarily add in quadrature to the total systematic uncertainty.
}
\label{tab:fit_0LSR}
\begin{tabular*}{\textwidth}{@{\extracolsep{\fill}}lccccc}
\noalign{\smallskip}\hline\noalign{\smallskip}
\textbf{b0L- Signal Region}           & SRA350            & SRA450            & SRA550    & SRB & SRC         \\[-0.05cm]
\noalign{\smallskip}\hline\noalign{\smallskip}
Observed          & $81$              & $24$              & $10$    &  45 &  7               \\
\noalign{\smallskip}\hline\noalign{\smallskip}
Total background (fit)     & $70 \pm 13$          & $22 \pm 5\phantom{0}$    & $7.2 \pm 1.5$   & $37 \pm 7\phantom{0}$   & $5.5 \pm 1.5$      \\
\noalign{\smallskip}\hline\noalign{\smallskip}
        $Z$+jets         & $46 \pm 12$          & $13.6 \pm 3.7\phantom{0}$          & $4.0 \pm 1.2$ & $20.0 \pm 5.2\phantom{0}$ & $2.3 \pm 0.8$        \\
        \ttbar\         & $2.0 \pm 0.6$          & $0.5 \pm 0.2$          & $0.16 \pm 0.07$   & $5.1 \pm 2.7$ &  $0.8 \pm 0.3$        \\
        Single top      & $4.7 \pm 3.4$          & $1.2 \pm 1.0$          & $0.5 \pm 0.3$   & $2.6 \pm 1.1$ & $0.7\pm 0.3$          \\
        $W$+jets        & $15  \pm 5\phantom{0}$          & $5.0 \pm 1.8$          & $2.4 \pm 1.0$  & $5.5 \pm 2.0$ & $1.3 \pm 0.8$          \\
        Others         & $2.5 \pm 1.7$          & $1.4 \pm 1.2$          & $0.07 \pm 0.03$ & $4.0 \pm 1.1$  & $0.4 \pm 0.1$      \\
 \noalign{\smallskip}\hline\noalign{\smallskip}
Total background (MC exp.)      & $60.4$          & $18.5$          & $6.2$  & $28$ &  $5.4$         \\
\noalign{\smallskip}\hline\noalign{\smallskip}
        $Z$+jets         & $34.9$          & $10.3$          & $3.0$  & $13.1$ & $1.9$          \\
        \ttbar\         & $1.9$          & $0.45$          & $0.16$ & $3.8$ &  $1.2$          \\
        Single top      & $10$          & $2.5$          & $1.0$  & $2.6$  & $0.7$         \\
        $W$+jets         & $11.6$          & $4.0$          & $1.9$  & $4.2$ & $1.2$          \\
        Others       & $2.5$          & $1.3$          & $0.07$  & $4.0$ &  $0.4$        \\
\noalign{\smallskip}\hline\noalign{\smallskip}
\end{tabular*}
}
\end{center}
\end{table}

\begin{table}
\begin{center}
\setlength{\tabcolsep}{0.0pc}
{\small
\caption{
Fit results in the b1L signal regions. The background normalization parameters are obtained
 from the background-only fit in the control regions and are applied to the SRs. 
Smaller backgrounds such as diboson, $Z$+jets, multijet and rare processes are indicated as ``Others''.   
The individual 
uncertainties, including detector-related and theoretical systematic components, are symmetrized and can be correlated. 
They do not necessarily add in quadrature to the total systematic uncertainty.
}
\label{tab:fit_1LSR}
\begin{tabular*}{\textwidth}{@{\extracolsep{\fill}}lcccc}
\noalign{\smallskip}\hline\noalign{\smallskip}
\textbf{b1L- Signal Region}                 & SRA600            & SRA750    & SRB & SRA300-2j      \\[-0.05cm]
\noalign{\smallskip}\hline\noalign{\smallskip}
Observed                      & $21$              & $13$    & 69  &  12               \\
\noalign{\smallskip}\hline\noalign{\smallskip}
Total background (fit)           & $24 \pm 6\phantom{0}$    & $15 \pm 4\phantom{0}$   & $53 \pm 12$   & $6.7 \pm 2.3$      \\
\noalign{\smallskip}\hline\noalign{\smallskip}
        \ttbar\                  & $10 \pm 5\phantom{0}$          & $5.5 \pm 2.7$   & $16 \pm 7\phantom{0}$ &  $2.4 \pm 1.3$        \\
        Single top                & $7 \pm 4$          & $4.5 \pm 2.8$   & $10 \pm 5\phantom{0}$ & $3.3\pm 2.0$          \\
        $W$+jets                 & $0.9 \pm 0.5$          & $0.6 \pm 0.3$  & $17 \pm 8\phantom{0}$ & $0.4 \pm 0.3$          \\
        $ttV$                    & $5.4 \pm 0.6$          & $4.0 \pm 0.5$ & $9 \pm 1$ & $0.6 \pm 0.1$        \\
        Others                & $0.07 \pm 0.02$          & $0.07 \pm 0.03$ & $1.8 \pm 0.3$  & $0.07 \pm 0.02$      \\
 \noalign{\smallskip}\hline\noalign{\smallskip}
Total background (MC exp.)      & $27$          & $17$  & $52$ &  $8.4$         \\
\noalign{\smallskip}\hline\noalign{\smallskip}
        \ttbar\                 & $9$          & $5.1$ & $16$ &  $2.2$          \\
        Single top              & $11$          & $7.1$  & $14$  & $5.2$         \\
        $W$+jets               & $0.9$          & $0.6$  & $11$ & $0.4$          \\
        $ttV$                & $5.4$          & $4.0$  & $9$ & $0.6$          \\
        Others              & $0.07$          & $0.07$ & $1.8$  & $0.07$      \\
\noalign{\smallskip}\hline\noalign{\smallskip}
\end{tabular*}
}
\end{center}
\end{table}

\begin{figure}[t]
\begin{center}
\includegraphics[width=0.99\textwidth]{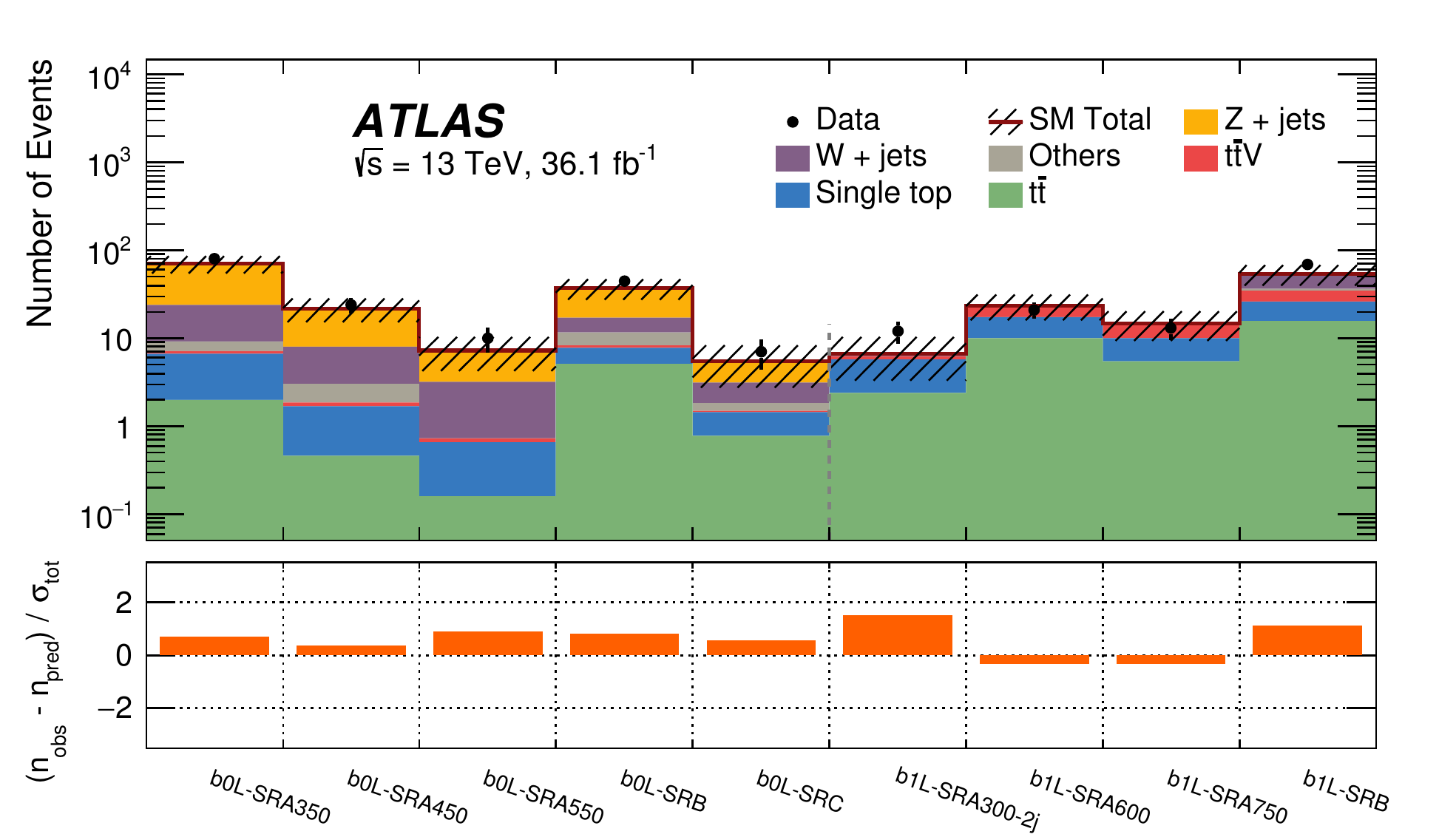}
\end{center}
\caption{ Results of the likelihood fit extrapolated to the SRs associated with the b0L and b1L analyses. 
The normalization of the backgrounds is obtained from the fit to the CRs. 
The upper panel shows the observed number of events and the predicted background yields. 
The contributions from diboson, multijet and rare backgrounds are collectively called ``Others''. 
All uncertainties defined in Section~\ref{sec:systematics}
 are included in the uncertainty band.  The lower panel shows the pulls in each SR, where $\sigma_{\textrm{tot}}$ is the error on the background estimation as a sum in quadrature of the systematic uncertainty and the statistical uncertainty on the estimate.}
\label{fig:SRpulls}
\end{figure}

Figure~\ref{fig:SRs} shows the comparison between the observed data 
and the SM predictions for some relevant kinematic distributions for the b0L and b1L selections. 
For illustrative purposes, the distributions expected for scenarios with different bottom 
squark and neutralino masses depending on the SR considered are shown.   

\begin{figure}[t]
\begin{center}
\vspace{-1.2cm}
\captionsetup[subfigure]{labelformat=simple}
\subfloat[\label{fig:SRA350_mctMbb}]{
\includegraphics[width=0.49\textwidth]{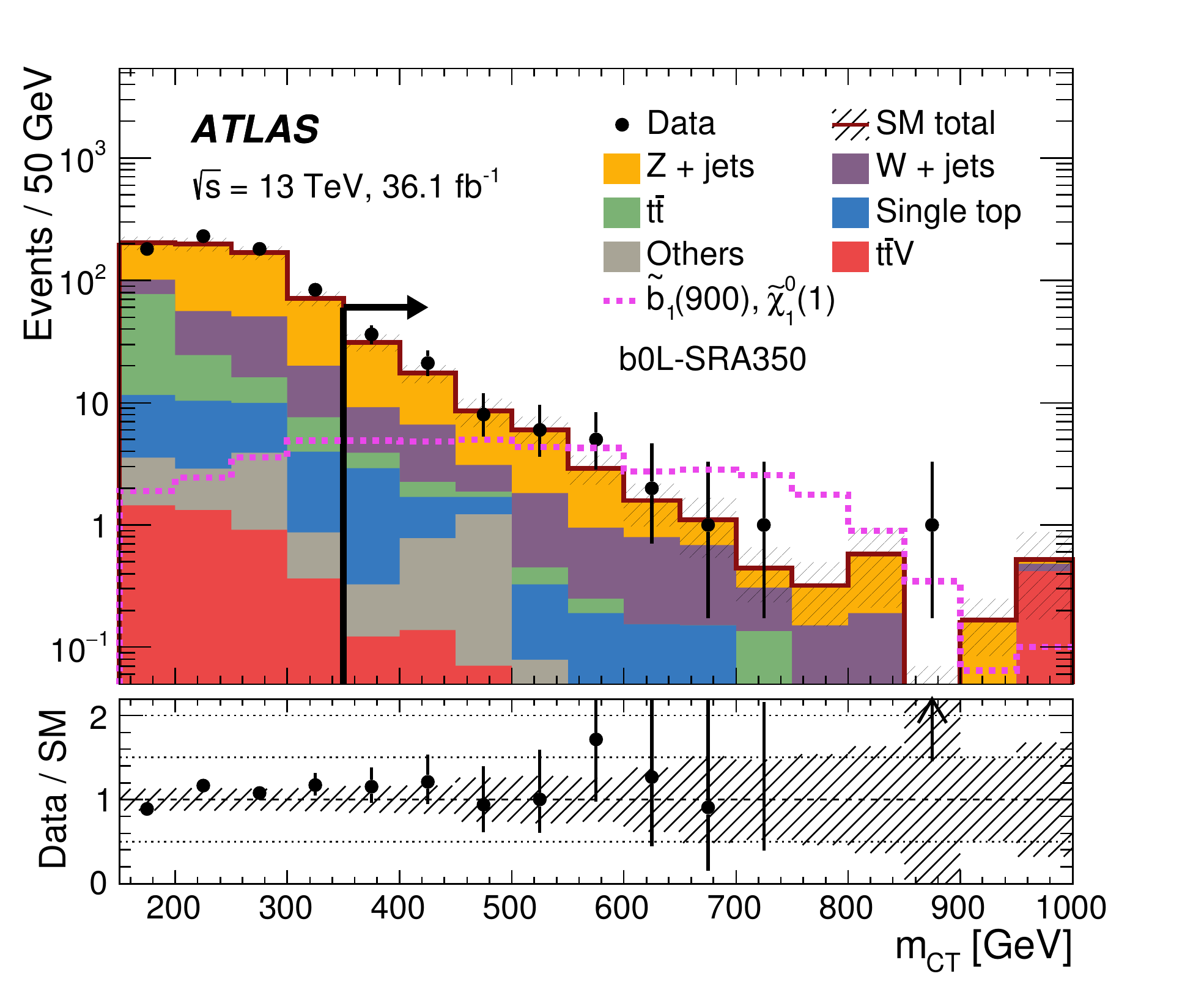}
}
\subfloat[\label{fig:SRB_mctETmiss}]{
\includegraphics[width=0.49\textwidth]{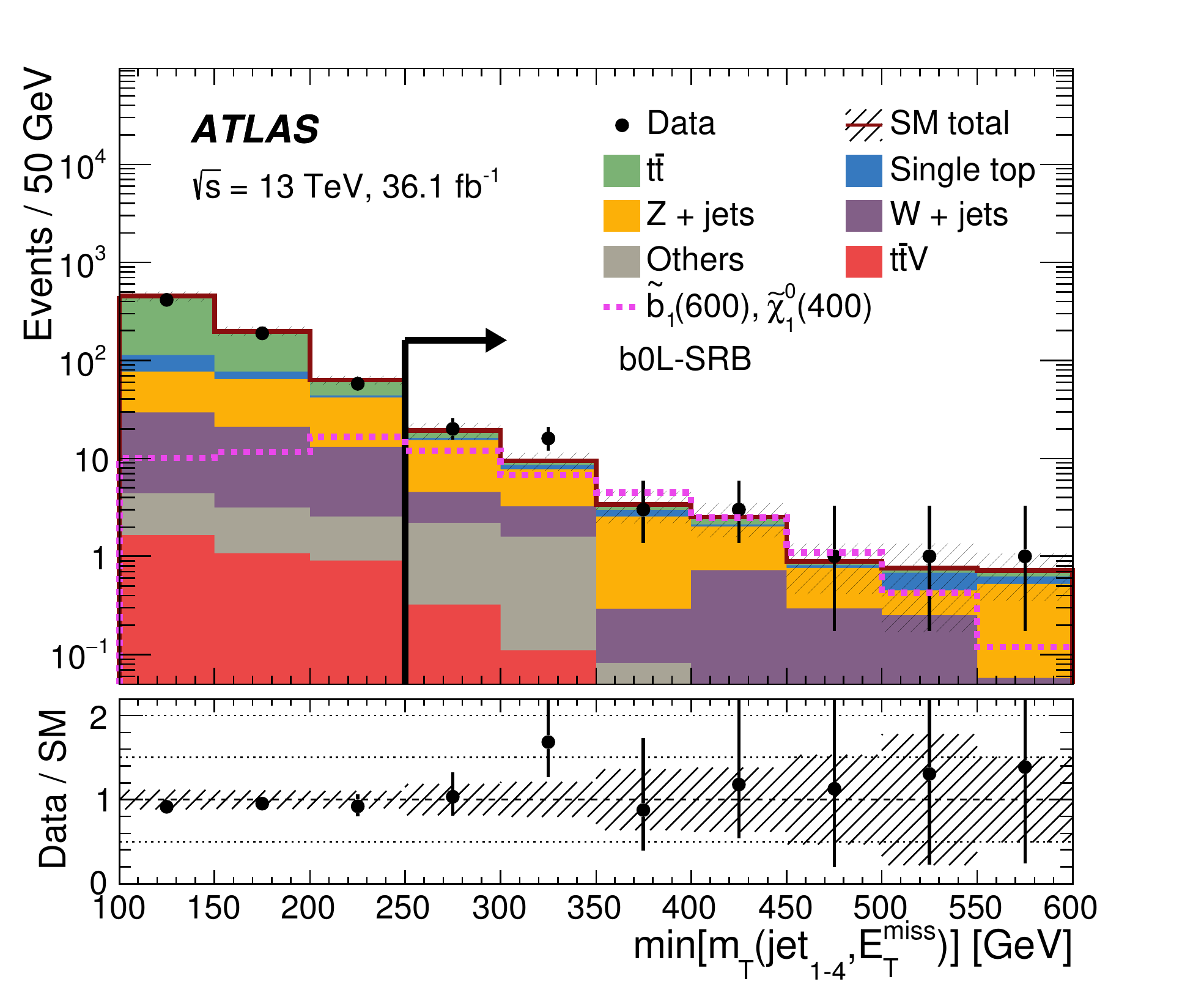}
} 
\vspace{-0.4cm}
\\
\subfloat[\label{fig:SRC_Asym}]{
\includegraphics[width=0.49\textwidth]{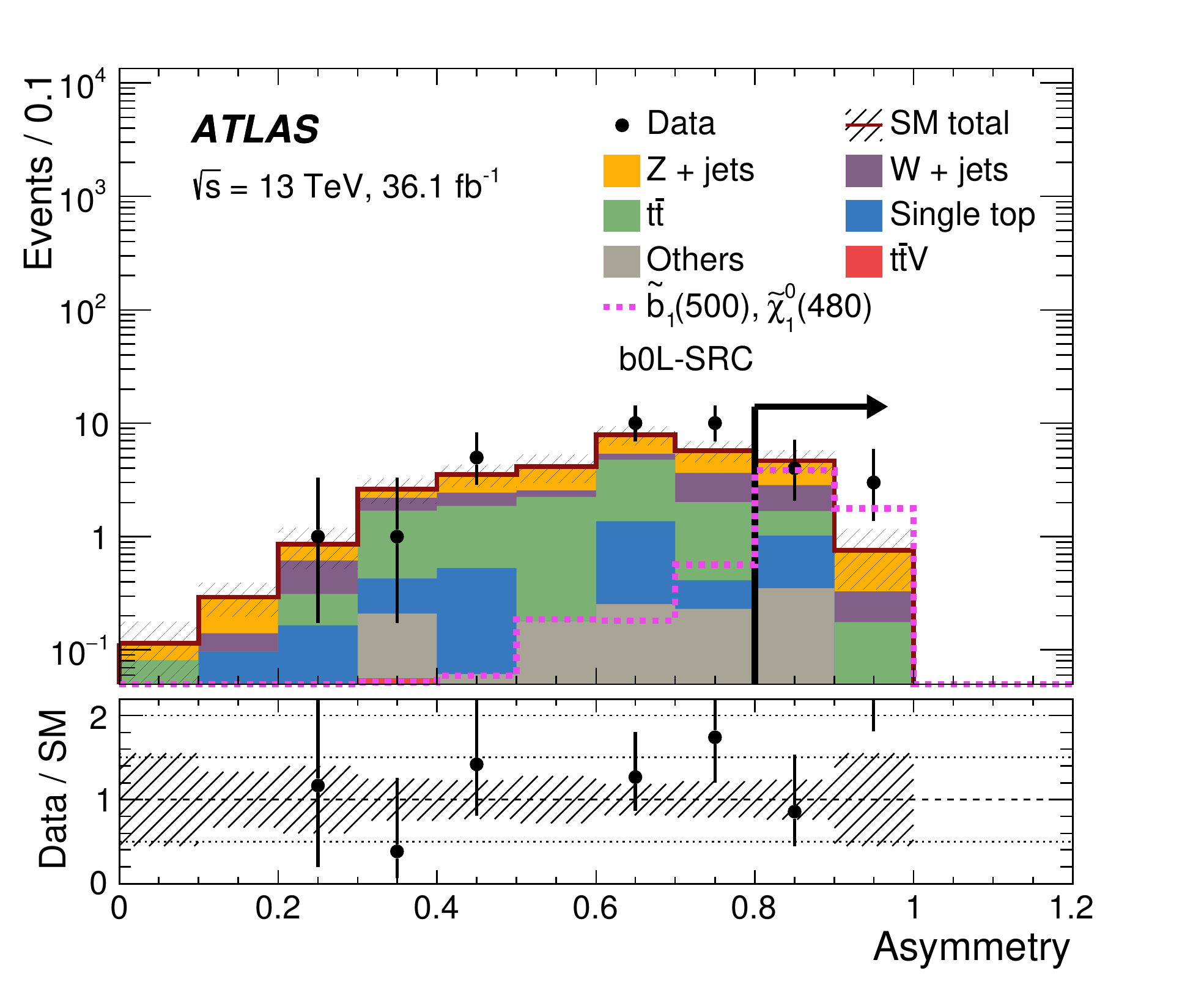}
} 
\subfloat[\label{fig:tbSRA3002j_mbb}]{
\includegraphics[width=0.49\textwidth]{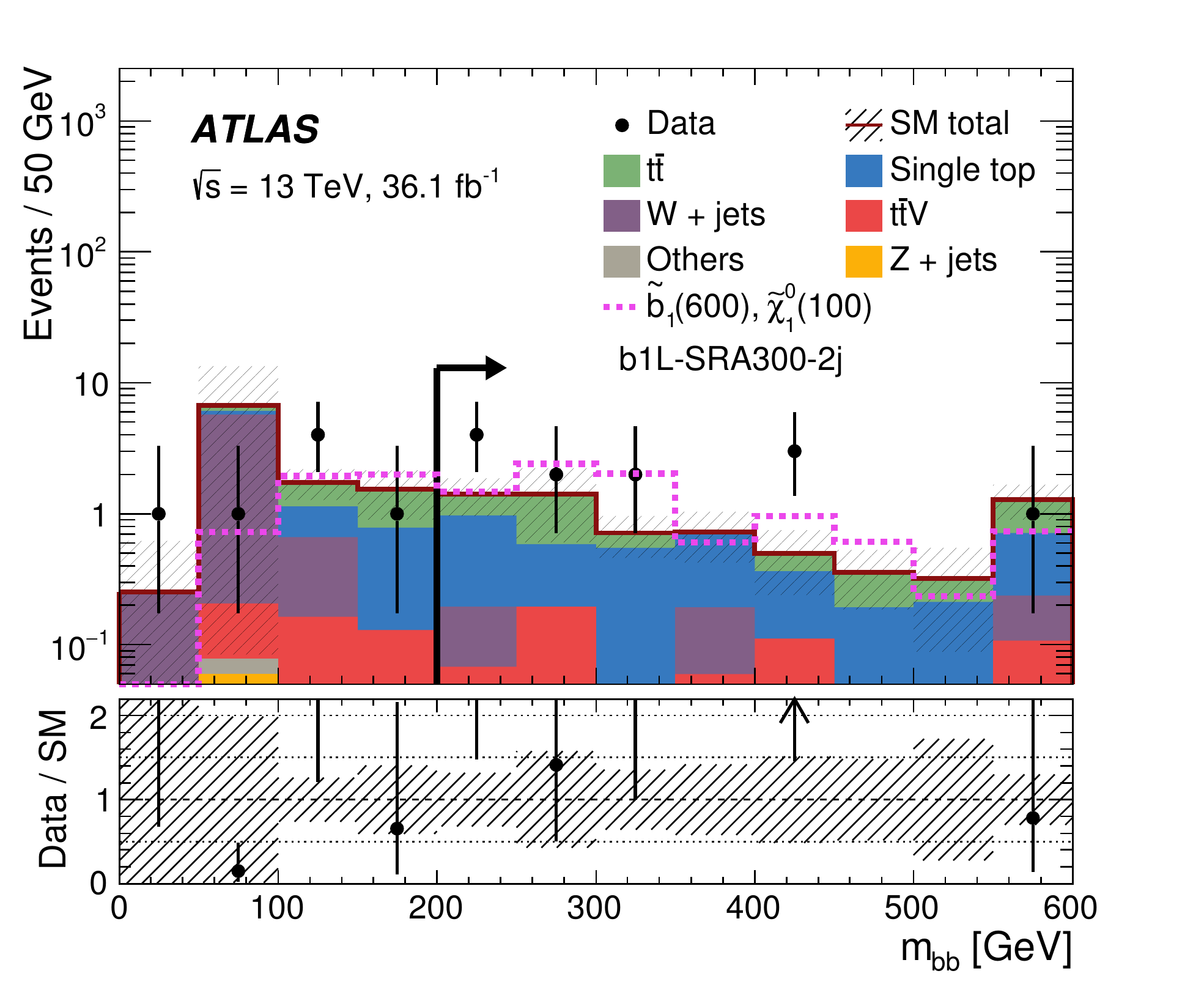}
}
\vspace{-0.4cm}
\\
\subfloat[\label{fig:tbSRA_mctMbb}]{
\includegraphics[width=0.49\textwidth]{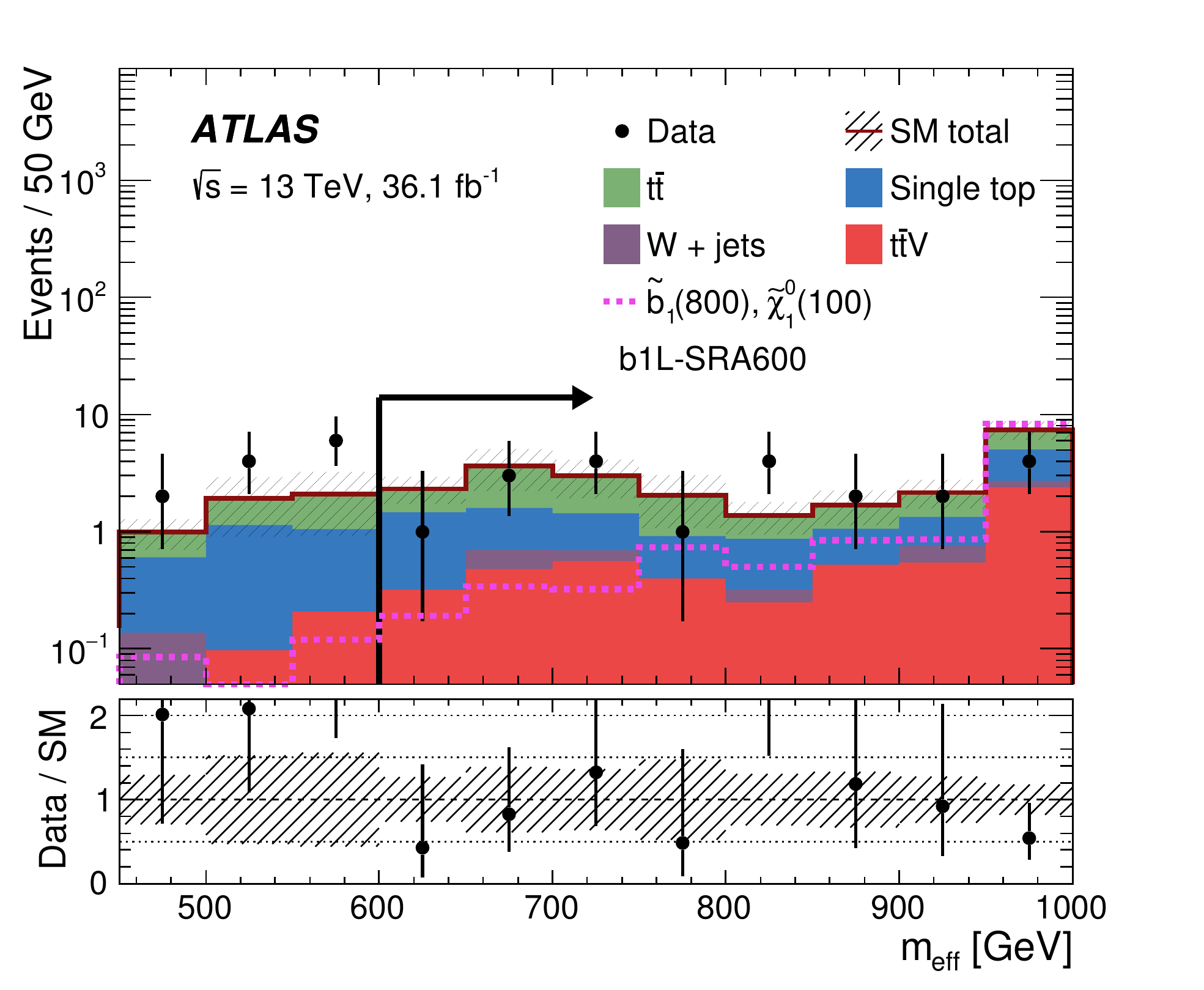}
}
\subfloat[\label{fig:tbSRB_mctETmiss}]{
\includegraphics[width=0.49\textwidth]{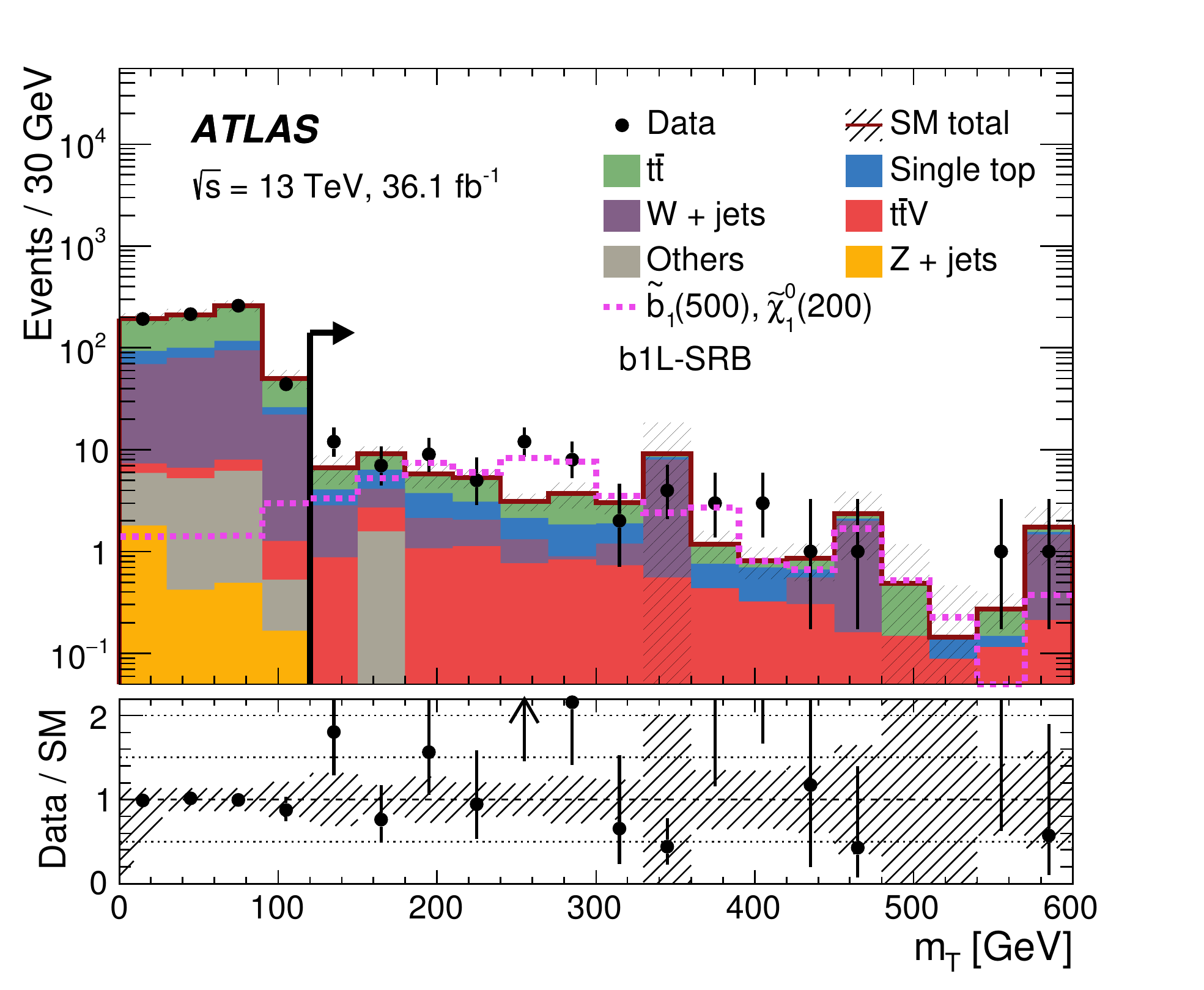}
} 
\vspace{-0.7cm}
\end{center}
\caption{ \label{fig:SRs} Distributions of (a) $\mct$ in b0L-SRA, (b) $\mtjmin$  in b0L-SRB,  
(c) ${\cal A}$  in b0L-SRC, (d) \mbb\  in b1L-SRA300-2j,
 (e) $\meff$  in b1L-SRA,  (f) $\mt$  in b1L-SRB.
All  selection criteria are applied, except the selection on the variable that is displayed in each of the plots. 
The arrows indicate the final selection applied in the signal regions. 
The shaded-grey band shows the detector-related systematic uncertainties and the statistical uncertainties on the MC samples.  
The SM backgrounds are normalized to the values determined in the fit. 
The contributions from diboson, multijet and rare backgrounds are collectively called ``Others''. 
For illustration the distribution expected from selected signal models are overlaid. The last bin includes overflow events. }

\end{figure}

The results are translated into upper limits on contributions from physics beyond the SM (BSM)
for each signal region.
The CL$_\mathrm{s}$~method~\cite{Junk:1999kv,Read:2002hq} is used to derive the confidence level
of the exclusion; signal models with a CL$_\mathrm{s}$~value below 0.05 are said to be excluded at 95\% CL.
The profile-likelihood-ratio test statistic is used to exclude the signal-plus-background hypothesis for
specific signal models.
$S^{95}_\mathrm{obs}$ ($S^{95}_\mathrm{exp}$) is the observed (expected) upper limit at 95\% CL on the number of events from BSM
phenomena for each signal region. These limits, when normalized by the integrated luminosity of the
data sample, may be interpreted as upper limits on the visible cross-section of BSM physics, $\sigma^{}_\textrm{vis}$,
 defined as the product of the production cross-section, the acceptance and the selection efficiency of a BSM signal. 
Table~\ref{tab:upper} summarizes $S^{95}_\mathrm{obs}$, $S^{95}_\mathrm{exp}$, and $\sigma^{}_\textrm{vis}$ for all SRs, together with
the $p_0$-values, which represent the probability of the SM background alone to fluctuate to the observed number of events or higher.
\begin{table}
\begin{center}
\caption[Breakdown of upper limits.]{
Left to right: 95\% CL upper limits on the visible cross-section
($\langle\epsilon A \sigma\rangle_\textrm{obs}^{95}$) and on the number of
signal events ($S_\textrm{obs}^{95}$ ).  The third column
($S_\textrm{exp}^{95}$) shows the 95\% CL upper limit on the number of
signal events, given the expected number (and $\pm 1\sigma$
variations of the expected number) of background events. The last column reports the $p_0$-values and $Z$ (the number of equivalent Gaussian standard deviations). The maximum allowed $p_0$-value is truncated at 0.5.
\label{tab:upper}}
\begin{tabular}{@{}lcccc@{}}
\toprule
\textbf{Signal channel}                        & $\langle\epsilon A \sigma\rangle_\textrm{obs}^{95}$[fb]  &  $S_\textrm{obs}^{95}$  & $S_\textrm{exp}^{95}$ & $p_0\ (Z)$\\
\midrule
b0L-SRA350   & $1.06$ &  $38.2$  & ${30.9}^{+11.3}_{-8.4}$  & 0.28 (0.60) \\[0.1cm]
b0L-SRA450   & $0.43$ &  $15.6$  & ${13.9}^{+5.6\phantom{0}}_{-3.8}$   & 0.37 (0.34)\\[0.1cm]
b0L-SRA550   & $0.30$ &  $10.7$  & ${\phantom{0}7.8}^{+3.7\phantom{0}}_{-1.6}$  &  0.20 (0.85)\\[0.1cm]
b0L-SRB      & $0.72$ &  $26.1$  & ${19.9}^{+8.3\phantom{0}}_{-5.4}$ & 0.23 (0.74)  \\[0.1cm]
b0L-SRC      & $0.24$ &  $8.7$   & ${\phantom{0}6.8}^{+3.3\phantom{0}}_{-1.3}$  & 0.30 (0.54) \\[0.1cm]
\hline
b1L-SRA300-2j & $0.39$ &  $14.1$   & ${\phantom{0}9.3}^{+3.5\phantom{0}}_{-3.1}$  & 0.08 (1.43) \\[0.1cm] 
b1L-SRA600    & $0.38$ &  $13.6$   & ${14.8}^{+5.4\phantom{0}}_{-4.4}$ & 0.50 (0.00)  \\[0.1cm] 
b1L-SRA750    & $0.27$ &  $9.9$   & ${11.2}^{+4.0\phantom{0}}_{-2.3}$  & 0.50 (0.00)  \\[0.1cm] 
b1L-SRB       & $1.12$ &  $40.3$   & ${28.7}^{+10.7}_{-8.2}$ & 0.21 (0.80)  \\[0.1cm] 
\bottomrule
\end{tabular}
\end{center}
\end{table}

Exclusion limits are obtained assuming two types of SUSY particle mass hierarchy such that 
the lightest bottom squark decays either exclusively via $\bone \rightarrow b \neut$ or 
into multiple channels, $\bone \rightarrow b \neut$ and $\bone \rightarrow t \chipm$, 
assuming a 50\% branching ratio and $\dmCN \sim$ 1~\GeV. The first set of scenarios is 
targeted by the zero-lepton channel SRs only. 
For models with mixed decays, the expected limits from the SRs are compared and the observed limits are obtained by statistically combining the most sensitive zero-lepton SR with the most sensitive one-lepton SR.
In all cases, the fit procedure takes into account correlations in the yield predictions between 
control and signal regions due to common background normalization parameters and systematic uncertainties.
The experimental systematic uncertainties in the signal are taken into account for this calculation 
and are assumed to be fully correlated with those in the SM background. 

For the exclusive $\bone \rightarrow b \neut$ decay mode, at each point of the parameter space the 
SR with the best expected sensitivity is used. 
Sensitivity to scenarios with the largest mass difference between the $\bone$ and the $\neut$~is achieved 
with the most stringent \mct~threshold (b0L-SRA550). Sensitivity to scenarios with intermediate and small  
mass differences is obtained with the dedicated b0L-SRB and b0L-SRC selections, respectively.  
For the mixed-decays scenarios, a statistical combination is computed with the results of the zero-lepton 
and one-lepton channels as explained above. 
A combined fit is performed simultaneously on the control and signal 
regions of the two analyses. The best sensitivity to regions of the ($\bone$, $\neut$) mass plane 
close to the kinematic boundaries is obtained with the combination of the b0L-SR450    
and b1L-SRB regions whilst stringent constraints on models with large mass differences are achieved 
with the results from the combination of zero-lepton and one-lepton SRs with the most 
stringent \mct~and \meff~thresholds, respectively. 

Figures~\ref{fig:bbLimit} and~\ref{fig:combLimit} show the observed (solid line) and expected (dashed line) 
exclusion contours at 95\% CL in the $\bone$--$\neut$ mass plane for the two types of SUSY scenarios 
considered.   
Bottom squark masses up to 950 (860)~\GeV\ are excluded for $\neut$~masses below 420 (250)~\GeV\ in 
models with exclusive (mixed) decay modes. Multiple-decay bottom squark models are phenomenologically 
equivalent to models characterized by the pair production of top squarks decaying 
as $\tone \rightarrow t \neut$ and $\tone \rightarrow b \chipm$, with the same assumptions for the 
branching ratios and  $\dmCN$. Hence the results can be interpreted as exclusion limits on 
top squark masses.

\begin{figure}[t]
\begin{center}
\subfloat[\label{fig:bbLimit}]{
\includegraphics[width=0.49\textwidth]{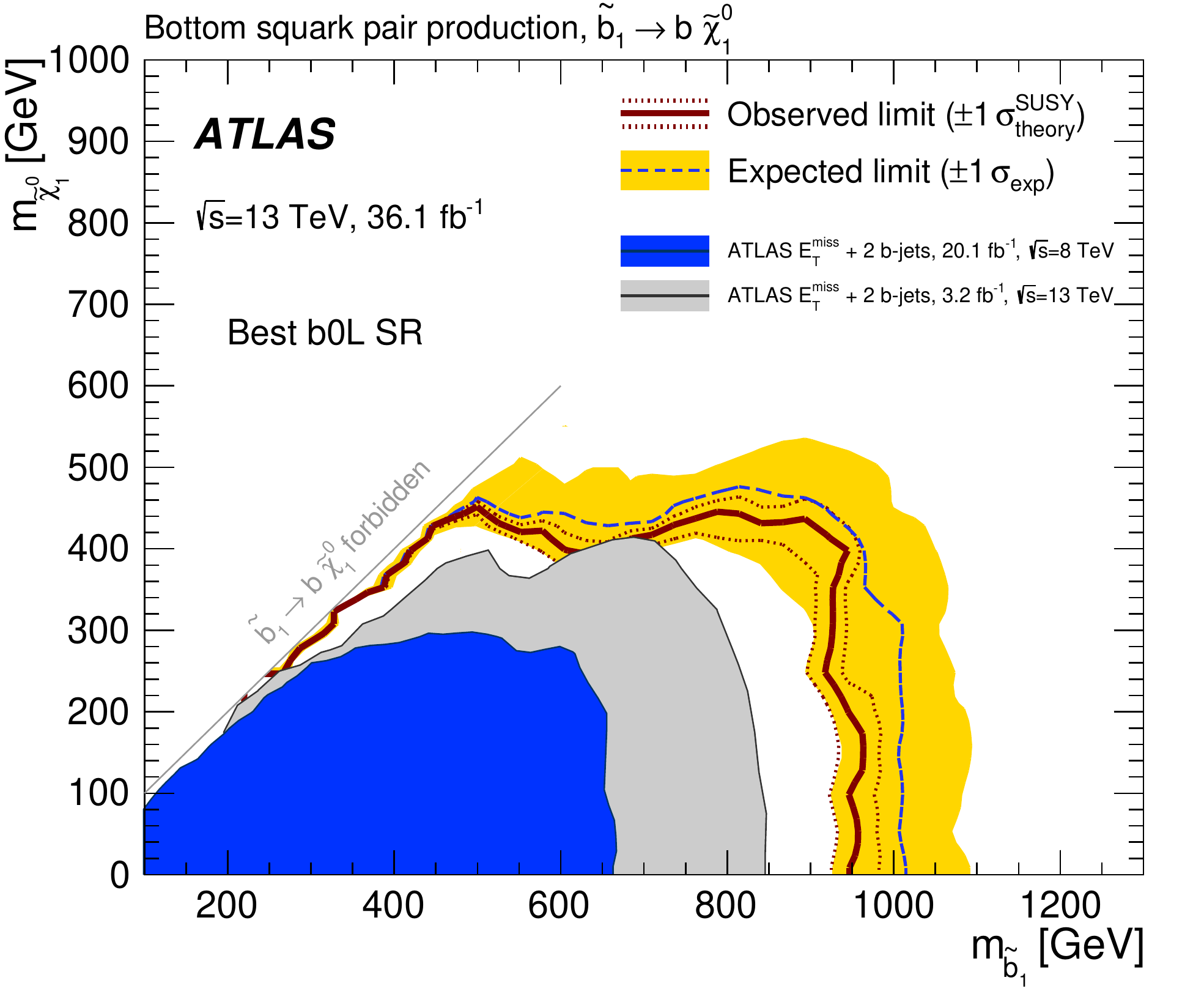}
}
\subfloat[\label{fig:combLimit}]{
\includegraphics[width=0.49\textwidth]{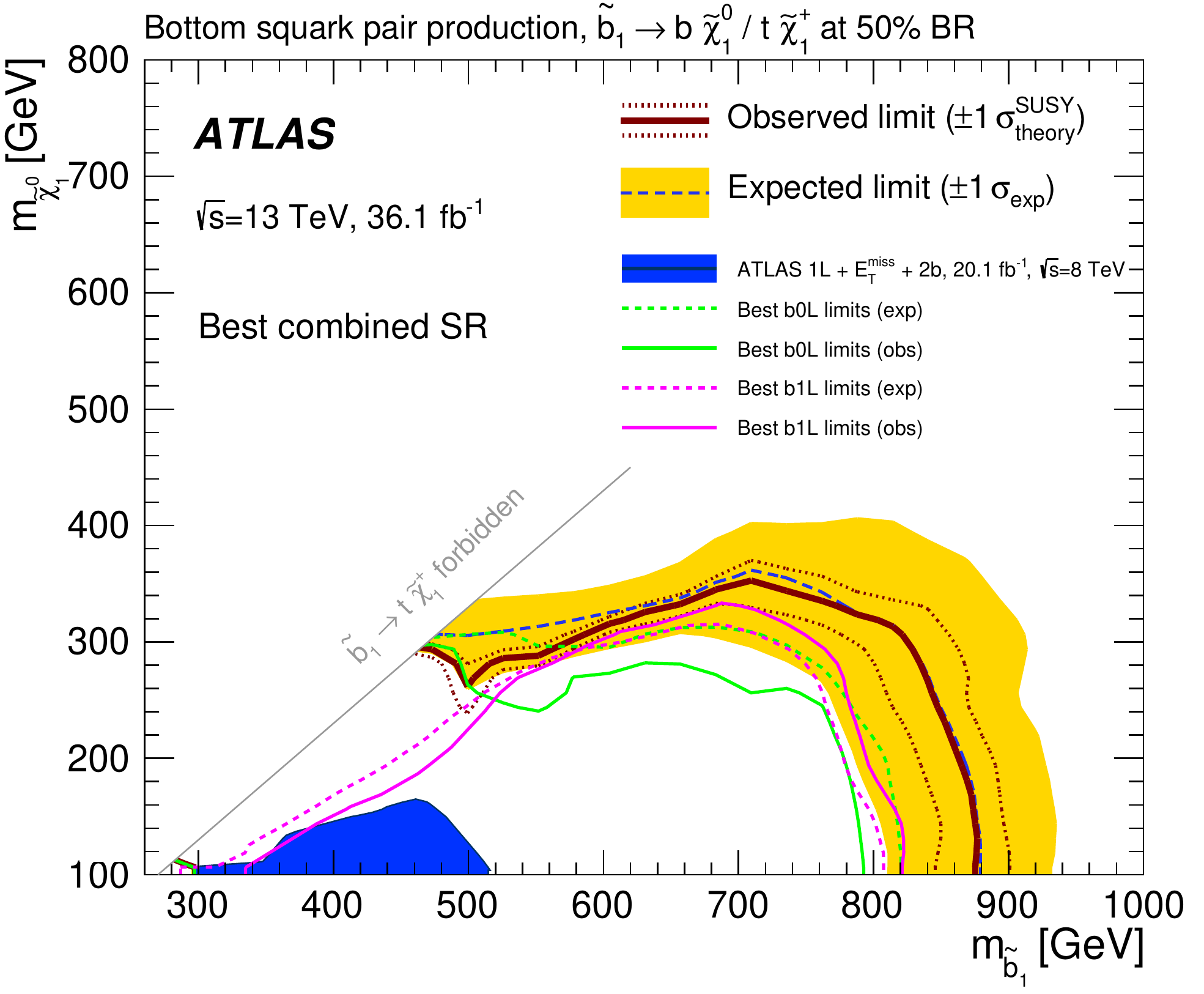}
}
\end{center}
\caption{ \label{fig:limits} (a) 
Observed and expected exclusion contours at 95\% CL, as well as $\pm 1 \sigma$ variation of the expected limit,  
in the $\bone$--$\neut$ mass plane. The SR with the best expected sensitivity is adopted for each point of the 
parameter space. The yellow band around the expected limit (dashed line) shows the impact of the 
experimental and SM background theoretical uncertainties. The dotted lines show the impact on the observed 
limit of the variation of the nominal signal cross-section by $\pm 1 \sigma$ of its theoretical uncertainties. 
(b) Same, for scenarios where multiple decay modes are considered for bottom squarks. The statistical 
combination of b0L-SRs and b1L-SRs is used in this case.}

\end{figure}

\FloatBarrier

\section{Conclusion}
\label{sec:conclusion}
The results of a search for pair production of bottom and top squarks are reported. The analysis uses \intlumi~of $pp$ 
collisions at $\sqrt{s}=13$~\TeV\ collected by the ATLAS experiment at the Large Hadron Collider in 2015 and 2016.
Third-generation squarks are searched for in events containing large missing transverse momentum and jets, exactly two 
of which are identified as \bjets. Selections are defined with either no charged leptons (electrons and
muons) in the final state, or one charged lepton. Zero-lepton channel signal regions target  
$R$-parity-conserving models in which the $\bone$ is the lightest squark and is assumed to decay exclusively
via $\bone \rightarrow b \neut$, where \neut~is the lightest neutralino. One-lepton channel signal regions 
target models where bottom or top squarks are produced and can decay into 
 multiple channels, 
$\bone \rightarrow b \neut$ and  $\bone \rightarrow t \chipm$, or $\tone \rightarrow t \neut$ and $\tone \rightarrow b \chipm$, where $\chipm$ is the lightest chargino and the mass difference $m_{\chipm}- m_{\neut}$ is set to 1~\GeV.  
No significant excess above the expected Standard Model background is found and exclusion limits at 95\% confidence 
level are placed on the visible cross-section and on the mass of the bottom (or top) squark. 
Bottom squark masses up to 950~\GeV\ are excluded for $\neut$~masses below 420~\GeV\ in
models with exclusive decay modes. Bottom or top squark masses up to 860~\GeV\ are excluded 
for $\neut$~masses below 250~\GeV\ in models with mixed decay modes with equal branching ratios.
The results significantly improve upon previous Run-1 and Run-2 searches at the ATLAS experiment  
and strengthen the constraints on bottom and top squark masses.

\section*{Acknowledgements}


We thank CERN for the very successful operation of the LHC, as well as the
support staff from our institutions without whom ATLAS could not be
operated efficiently.

We acknowledge the support of ANPCyT, Argentina; YerPhI, Armenia; ARC, Australia; BMWFW and FWF, Austria; ANAS, Azerbaijan; SSTC, Belarus; CNPq and FAPESP, Brazil; NSERC, NRC and CFI, Canada; CERN; CONICYT, Chile; CAS, MOST and NSFC, China; COLCIENCIAS, Colombia; MSMT CR, MPO CR and VSC CR, Czech Republic; DNRF and DNSRC, Denmark; IN2P3-CNRS, CEA-DSM/IRFU, France; SRNSF, Georgia; BMBF, HGF, and MPG, Germany; GSRT, Greece; RGC, Hong Kong SAR, China; ISF, I-CORE and Benoziyo Center, Israel; INFN, Italy; MEXT and JSPS, Japan; CNRST, Morocco; NWO, Netherlands; RCN, Norway; MNiSW and NCN, Poland; FCT, Portugal; MNE/IFA, Romania; MES of Russia and NRC KI, Russian Federation; JINR; MESTD, Serbia; MSSR, Slovakia; ARRS and MIZ\v{S}, Slovenia; DST/NRF, South Africa; MINECO, Spain; SRC and Wallenberg Foundation, Sweden; SERI, SNSF and Cantons of Bern and Geneva, Switzerland; MOST, Taiwan; TAEK, Turkey; STFC, United Kingdom; DOE and NSF, United States of America. In addition, individual groups and members have received support from BCKDF, the Canada Council, CANARIE, CRC, Compute Canada, FQRNT, and the Ontario Innovation Trust, Canada; EPLANET, ERC, ERDF, FP7, Horizon 2020 and Marie Sk{\l}odowska-Curie Actions, European Union; Investissements d'Avenir Labex and Idex, ANR, R{\'e}gion Auvergne and Fondation Partager le Savoir, France; DFG and AvH Foundation, Germany; Herakleitos, Thales and Aristeia programmes co-financed by EU-ESF and the Greek NSRF; BSF, GIF and Minerva, Israel; BRF, Norway; CERCA Programme Generalitat de Catalunya, Generalitat Valenciana, Spain; the Royal Society and Leverhulme Trust, United Kingdom.

The crucial computing support from all WLCG partners is acknowledged gratefully, in particular from CERN, the ATLAS Tier-1 facilities at TRIUMF (Canada), NDGF (Denmark, Norway, Sweden), CC-IN2P3 (France), KIT/GridKA (Germany), INFN-CNAF (Italy), NL-T1 (Netherlands), PIC (Spain), ASGC (Taiwan), RAL (UK) and BNL (USA), the Tier-2 facilities worldwide and large non-WLCG resource providers. Major contributors of computing resources are listed in Ref.~\cite{ATL-GEN-PUB-2016-002}.



\bibliographystyle{bibtex/bst/atlasBibStyleWithTitle}
\bibliography{SbottomPaper2017,susy_v3,bibtex/bib/ATLAS,bibtex/bib/CMS,bibtex/bib/ConfNotes,bibtex/bib/PubNotes}
\newpage 
\begin{flushleft}
{\Large The ATLAS Collaboration}

\bigskip

M.~Aaboud$^\textrm{\scriptsize 137d}$,
G.~Aad$^\textrm{\scriptsize 88}$,
B.~Abbott$^\textrm{\scriptsize 115}$,
O.~Abdinov$^\textrm{\scriptsize 12}$$^{,*}$,
B.~Abeloos$^\textrm{\scriptsize 119}$,
S.H.~Abidi$^\textrm{\scriptsize 161}$,
O.S.~AbouZeid$^\textrm{\scriptsize 139}$,
N.L.~Abraham$^\textrm{\scriptsize 151}$,
H.~Abramowicz$^\textrm{\scriptsize 155}$,
H.~Abreu$^\textrm{\scriptsize 154}$
% ...
\bigskip
\\
% ...
$^{12}$ Institute of Physics, Azerbaijan Academy of Sciences, Baku, Azerbaijan\\
% ...
$^{88}$ CPPM, Aix-Marseille Universit{\'e} and CNRS/IN2P3, Marseille, France\\
% ...
$^{115}$ Homer L. Dodge Department of Physics and Astronomy, University of Oklahoma, Norman OK, United States of America\\
% ...
$^{119}$ LAL, Univ. Paris-Sud, CNRS/IN2P3, Universit{\'e} Paris-Saclay, Orsay, France\\
% ...
$^{137}$ $^{(a)}$ Facult{\'e} des Sciences Ain Chock, R{\'e}seau Universitaire de Physique des Hautes Energies - Universit{\'e} Hassan II, Casablanca; $^{(b)}$ Centre National de l'Energie des Sciences Techniques Nucleaires, Rabat; $^{(c)}$ Facult{\'e} des Sciences Semlalia, Universit{\'e} Cadi Ayyad, LPHEA-Marrakech; $^{(d)}$ Facult{\'e} des Sciences, Universit{\'e} Mohamed Premier and LPTPM, Oujda; $^{(e)}$ Facult{\'e} des sciences, Universit{\'e} Mohammed V, Rabat, Morocco\\
% ...
$^{139}$ Santa Cruz Institute for Particle Physics, University of California Santa Cruz, Santa Cruz CA, United States of America\\
% ...
$^{151}$ Department of Physics and Astronomy, University of Sussex, Brighton, United Kingdom\\
% ...
$^{154}$ Department of Physics, Technion: Israel Institute of Technology, Haifa, Israel\\
$^{155}$ Raymond and Beverly Sackler School of Physics and Astronomy, Tel Aviv University, Tel Aviv, Israel\\
% ...
$^{161}$ Department of Physics, University of Toronto, Toronto ON, Canada\\
% ...
$^{a}$ Also at Department of Physics, King's College London, London, United Kingdom\\
$^{b}$ Also at Institute of Physics, Azerbaijan Academy of Sciences, Baku, Azerbaijan\\
$^{c}$ Also at Novosibirsk State University, Novosibirsk, Russia\\
$^{d}$ Also at TRIUMF, Vancouver BC, Canada\\
$^{e}$ Also at Department of Physics {\&} Astronomy, University of Louisville, Louisville, KY, United States of America\\
% ...
$^{*}$ Deceased
\end{flushleft}


%
%

\end{document}